\documentclass{emulateapj}
\usepackage{graphicx}
\usepackage{apjfonts}

\newcommand{\kmsend}{\mbox{km s$^{-1}$}}
 \newcommand{\kms}{\mbox{km s$^{-1}$ }}

\newcommand{\msun}{\mbox{M$_{\sun}$ }}

\newcommand{\lsunend}{\mbox{L$_{\sun}$}}
\newcommand{\lir}{\mbox{L$_{\rm IR}$}}

\newcommand{\lhcnjone}{\mbox{HCN (J=1-0)}}
\newcommand{\lco}{\mbox{L$_{\rm CO}$}}
\newcommand{\lmol}{\mbox{L$_{\rm mol}$}}
\newcommand{\lcojone}{\mbox{CO (J=1-0)}}
\newcommand{\lcojthree}{\mbox{CO (J=3-2)}}

\newcommand{\cmthree}{\mbox{cm$^{-3}$}}
\newcommand{\cmtwo}{\mbox{cm$^{-2}$}}

\newcommand{\msunyrend}{\mbox{M$_{\sun}$yr$^{-1}$}}
\newcommand{\htwo}{\mbox{H$_2$}}
\newcommand{\mhtwo}{\mbox{$M_{H2}$}}
\newcommand{\z}{\mbox{$z$}}

\newcommand{\zsim}{\mbox{$z\sim$ }}

\newcommand{\magorrian}{\mbox{$M_{\rm BH}$-$M_{\rm bulge}$ }}

\shorttitle{Molecular SFR Indicators in Galaxies}
\shortauthors{Narayanan et al.}

\slugcomment{Submitted to ApJ, November 8th, 2007}

\begin{document}
\title{Molecular Star Formation Rate Indicators in Galaxies}
\author{Desika Narayanan\altaffilmark{1}, Thomas
J. Cox\altaffilmark{2,3}, Yancy Shirley\altaffilmark{1,4}, Romeel
Dav\'e\altaffilmark{1}, Lars Hernquist\altaffilmark{2}, Christopher
K. Walker\altaffilmark{1}}

\altaffiltext{1}{Steward Observatory, University of Arizona, 933 N
Cherry Ave, Tucson, AZ, 85721, USA}

\altaffiltext{2}{Harvard-Smithsonian Center for Astrophysics, 60
Garden Street, Cambridge, MA 02138, USA} 
\altaffiltext{3}{W.M. Keck Fellow}
\altaffiltext{4}{Bart J. Bok Fellow}

\begin{abstract}
We derive a physical model for the observed relations between star
formation rate (SFR) and molecular line (CO and HCN) emission in
galaxies, and show how these observed relations are reflective of the
underlying star formation law. We do this by combining 3D non-LTE
radiative transfer calculations with hydrodynamic simulations of
isolated disk galaxies and galaxy mergers. We demonstrate that the
observed SFR-molecular line relations are driven by the relationship
between molecular line emission and gas density, and anchored by the
index of the underlying Schmidt law controlling the SFR in the
galaxy. Lines with low critical densities (e.g. CO J=1-0) are
typically thermalized and trace the gas density faithfully. In these
cases, the SFR will be related to line luminosity with an index
similar to the Schmidt law index. Lines with high critical densities
greater than the mean density of most of the emitting clouds in a
galaxy (e.g. CO J=3-2, HCN J=1-0) will have only a small amount of
thermalized gas, and consequently a superlinear relationship between
molecular line luminosity (\lmol) and mean gas density
($\bar{n}$). This results in a SFR-line luminosity index less than the
Schmidt index for high critical density tracers. One observational
consequence of this is a significant redistribution of light from the
small pockets of dense, thermalized gas to diffuse gas along the line
of sight, and prodigious emission from subthermally excited gas. At
the highest star formation rates, the SFR-\lmol \ slope tends to the
Schmidt index, regardless of the molecular transition.  The
fundamental relation is the Kennicutt-Schmidt law, rather than the
relation between SFR and molecular line luminosity. Our model for
SFR-molecular line relations quantitatively reproduces the slopes of
the observed SFR-CO (J=1-0), CO (J=3-2) and HCN (J=1-0) relations when
a Schmidt law with index of $\sim$1.5 describes the SFR.  We use these
results to make imminently testable predictions for the SFR-molecular
line relations of unobserved transitions.

\end{abstract}

\section{Introduction}

The rate at which stars form in galaxies has historically been
parameterized in terms of ``laws'' relating the star formation rate
(SFR) to the density of available gas. \citet{sch59} originally
proposed a power-law form for the SFR such that
SFR $\propto$ $\rho^N$ (hereafter, referred to as a Schmidt Law).

Observed SFR relations typically come in two flavors. The first, relating
surface SFR density to surface gas density takes the form:
\begin{equation}
\label{eq:ken}
\Sigma_{\rm SFR} \propto \Sigma_{\rm gas}^N \, .
\end{equation}
Observations of local galaxies have constrained the surface density
SFR index, $N$, to roughly $N$=1.4$\pm$0.15 \citep[e.g.][and references
therein]{ken98a,ken98b,ken07}.

The second varietal of SFR indicators relates the SFR to the mass of
molecular gas above a given volumetric density. For example, local
galaxy surveys have shown a relation exists between the SFR and
molecular gas such that the SFR (as traced by the infrared
luminosity - hereafter \lir) is proportional to the $^{12}$CO (J=1-0)
luminosity to the 1.4-1.6 power \citep[][and references
therein]{san91,sanmir96}. Because the J=1-0 transition of $^{12}$CO
(hereafter CO) can be excited at relatively low densities ($n
\sim$10$^2$-10$^{3}$ \cmthree), and lies a modest $\sim$5 K above
ground, it serves as a valuable tracer of total molecular \htwo \ gas
content down to relatively low densities. These observed relations
have been broadly interpreted as an increasing star formation
efficiency (SFE; the SFR divided by \mhtwo) as a function of molecular
gas mass \citep{gao04a,gao04b,san91}.

Recent observations of HCN have suggested that a perhaps more fundamental
volumetric SFR relation for galaxies exists in terms of the {\it
dense} molecular gas. Because HCN (J=1-0) has a relatively high
critical density ($n_{\rm crit} \sim$10$^5$ \cmthree) compared to that
of CO (J=1-0), HCN is typically only thermalized in the dense cores of
molecular clouds. Thus, in the limit that the bulk of HCN luminosity
originates from thermalized gas, HCN serves as a good tracer of the
dense molecular gas that is actively involved in the star formation
process. This is in contrast to CO (J=1-0) which tends to emit from
both dense cores as well as diffuse molecular filaments and cloud
atmospheres. 

Pioneering millimeter-wave observations by
\citet{gao04a,gao04b} uncovered a tight linear correlation between the
infrared luminosity and HCN luminosity in a sample of local galaxies,
ranging from quiescent spirals to luminous and ultraluminous infrared
galaxies (LIRGs and ULIRGs, respectively). This relationship was found
by \citet{wu05} to hold in individual star forming cores in the Milky
Way as well.  Observations of a roughly linear correlation between
\lir \ and CO (J=3-2) emission (with critical density $n_{\rm crit}
\sim$10$^4$ \cmthree) in a similar sample of galaxies further
corroborated this result, and provided evidence against HCN-related
chemistry driving the observed \citet{gao04a,gao04b} relations
\citep{nar05,yao03}.

Despite a plethora of observations, a consensus physical
interpretation of the seemingly disparate \lir-\lhcnjone,
\lir-\lcojone \ and \lir-\lcojthree \ relations has yet to be borne
out of the literature. \citet{gao04a,gao04b} suggest that the linear
relationship between \lir-\lhcnjone \ reflects a scenario in which the SFR
in galaxies is directly proportional to the amount of dense molecular
gas, and similarly, a constant star formation efficiency in terms of
dense molecular gas. In this picture, LIRGs and ULIRGs simply have a
higher fraction of their molecular gas in a dense gas phase which is
reflected in the linear \lir-\lhcnjone \ slope\footnote{Throughout
this paper we use the term slope interchangeably with 'index' (as in
the exponent in a given SFR relation - e.g. Equation~\ref{eq:ken}). We
do this as we are considering all relations in log-log space.}  and
non-linear \lir-\lcojone \ slope. \citet{nar05} suggest a similar
interpretation for their observed relation of \lir $\propto$
\lcojthree$^{0.92}$.

\citet{wu05} extended these interpretations to include the observed
linear \lir-\lhcnjone \ relation within the Galaxy. These observations
suggested that dense molecular cores may represent a fundamental unit
of star formation. In this case, the only difference between a ULIRG
like Arp 220 and a Galactic star forming region is the number of
individual star forming units present. This leads to a natural linear
relationship between the SFR and amount of dense molecular gas.

From the theoretical side, \citet{krumthom07} made headway towards
understanding the observed relations by providing a model which
quantitatively reproduces the observed \lir-\lcojone \ and
\lir-\lhcnjone \ relations for star forming clouds. By combining 1D
non-local thermodynamic equilibrium (LTE) radiative transfer
calculations with models of giant molecular clouds (GMCs), these
authors found that the relationship between the SFR and molecular line
luminosity in star forming clouds depends on how the critical density
of the molecule compares to the mean density of the observed
source. Lines with critical densities which are typically below the
mean density in a galaxy (e.g. CO J=1-0) probe the total molecular gas
commonly from galaxy to galaxy. In this scenario, the SFR-molecular
line exponent can be represented by the index that relates SFR to
total volumetric gas density (i.e. the Schmidt index).  Conversely,
observations of molecules which typically trace densities well above
the mean density of the galaxy (e.g. HCN J=1-0) trace similar
conditions from galaxy to galaxy - i.e. the peaks in the density
spectrum. In these cases, the molecular line luminosity rises faster
than linearly with increasing gas density and the corresponding
relation between SFR and line luminosity is near linear. A key
direction forward beyond these models is understanding the effects of
emitting GMCs on a galaxy-wide scale, with the potential effects of
molecular line radiative transfer.

In this paper, we build on the body of observational and theoretical
work by providing a physical model for the origin of the observed
molecular SFR indicators on galaxy-wide scales, and relating them to
observed relations. We additionally make model-distinguishing testable
predictions for how the SFR in galaxies relates to as yet unobserved
transitions in CO and HCN. We do this by combining 3D non-LTE
molecular line radiative transfer codes
\citep{nar06a,nar06b,nar07a,nar07b} with smoothed particle
hydrodynamic (SPH) simulations of both isolated star forming galaxies
and galaxy mergers. Our methodology includes the effects of both
collisional and radiative molecular excitation and de-excitation, a
multi-phase ISM, and star formation \citep{spr05a}. We additionally
include a methodology for black hole growth and the winds associated
with both black hole growth and star formation, though note that these
processes play a minimal role on the SFR-\lmol \ relation in our
simulations.

The paper is outlined as follows. In \S~\ref{section:methods} we
describe our hydrodynamic and radiative transfer simulations. In
\S~\ref{section:origins} we quantitatively describe the origin for the
observed SFR-CO and SFR-HCN slopes and follow by making testable
predictions for submillimeter-wave telescopes in
\S~\ref{section:predictions}. In \S~\ref{section:observations} we
compare our results to the body of observational work in this field,
and in \S~\ref{section:othertheory}, relate these simulations to other
models and interpretations for molecular SFR indicators. In
\S~\ref{section:conclusions}, we conclude with a summary.  Throughout
the work we assume a $\Lambda$CDM cosmology with $h$=0.7,
$\Omega_\Lambda$=0.7, $\Omega_M$=0.3.

\section{Numerical Methods}
\label{section:methods}

\subsection{Hydrodynamics}
\label{section:hydrodynamics}

For this work we have modeled both isolated disk galaxies and major
mergers. We do this as the \citet{gao04a,gao04b} and \citet{nar05}
samples include both disk galaxies as well as ongoing mergers. The
merger snapshots are spaced equally temporally with a sampling of 5
Myr. We note, however, that the results in this paper are not
dependent on our usage of any particular combination of disk galaxies
or mergers.

The hydrodynamic simulations were conducted with a modified version of
the publicly available $N$-body/SPH code GADGET-2 \citep{spr05b}.  The
prescriptions used to generate the galaxies, as well as the algorithms
involved in simulating the physics of the multi-phase ISM, star
formation, and black hole growth are described fully in in
\citet{spr05a} and \citet{sh02,sh03}. We refer the reader to these
works for further details though summarize the aspects most relevant
to this study here.

GADGET-2 accounts for radiative cooling of the gas
\citep{dav99,kat96}, and a multi-phase ISM which is considered to
consist of cold clouds embedded in a hot, pressure confining ISM
\citep[e.g.][]{mckost77}. This is realized numerically through
'hybrid' SPH particles in which cold clouds are allowed to grow
through radiative cooling of the hot ISM, and conversely star
formation can evaporate the cold medium into diffuse, hot
gas. Pressure feedback from supernovae heating is treated via an
effective equation of state (EOS) \citep[see Figure 4 of][]{spr05a}.
Here we utilize an EOS softening parameter of $q_{\rm EOS}$=0.25.

Star formation is constrained to fit the observed Kennicutt-Schmidt
laws \citep{ken98a,ken98b,sch59}. In particular, star formation is
assumed to be related to the gas density and local star formation time
scale:
\begin{equation}
\frac{d\rho_\star}{dt} = (1-\beta)\frac{\rho_{\rm cold}}{t_{\star}}
\label{eq:sfr1}
\end{equation}
where $\beta$ quantifies the mass fraction of stars which are
short-lived and effectively immediatley supernova, and $t_{\star}$ is
the timescale for star formation.

A Salpeter IMF with slope -1.35 and mass range 0.1-40 $\msun$ returns a
value of $\beta \sim$0.1.  The parameter $t_\star$ is assumed to be
proportional to the local dynamical time scale.
\begin{equation}
t_\star({\rho})=t^\star_0\left(\frac{\rho}{\rho_{\rm th}}\right)^{-1/2}
\end{equation}
Detailed studies by \cite{cox06c} and \citet{sh03} have found that a
proportionality constant of $t^\star_0$ = 2.1 Gyr reproduces both the
normalization and slope of the \citet{ken98a,ken98b} observed SFR
surface density relations over mergers with a wide variety of isolated
galaxy and galaxy merger models. The dispersion in the modeled
relations lay well within the observed dispersion. Similar results
have been found by \citet{mihos94a,mihos94b,mihos96} and \citet{spr00}.
Hence, we use this formulation for the star formation rate.

Black holes are optionally included in our simulations which accrete
via a Bondi-Lyttleton-Hoyle parameterization with a fixed maximum rate
corresponding to the Eddington limit. The black hole radiates such
that its bolometric luminosity is set by the accretion rate with
$L=\epsilon \dot{M} c^2$ with accretion efficiency $\epsilon$=0.1. We
further assume that 5\% of this energy couples isotropically to the
surrounding ISM as feedback energy \citep{dim05,spr05a}, a number
chosen to match the normalization of the locally observed \magorrian
relation. We have run identical simulations both with and without
black holes for this work, and found that the inclusion of black holes
makes little difference on the final result.  We note, however, that
we focus our molecular line comparisons to the SFR, and not \lir \
(which could, in principle, have a contribution from buried AGNs at
the highest luminosities) as a complete treatment of modeling the
infrared SED \citep[e.g.][]{cha07a,cha07b,jon06,li07b} is outside the
scope of this study.


We consider both isolated disk galaxies with varying gas fractions and
masses, as well as two equal mass binary mergers. We utilized 15
isolated galaxies exploring a parameter space with gas fractions=[0.2,
0.4, 0.8] and virial velocities=[115, 160, 225, 320, 500 \kmsend]. All
disk galaxies (including the progenitors for the merger simulations)
were initialized with a \citet{her90} 
dark matter profile, concentration index
$c$=9, and spin parameter $\lambda$=0.033. For all galaxies we utilize
120,000 dark matter particles, and 80,000 total disk particles. The
softening lengths were 100 pc for baryons and 200 pc for dark
matter.

For the mergers, the progenitor galaxies were identical to the disk
galaxies described here, with $V_{\rm 200}$=160 \kmsend, and 40\% gas
fraction. For reference, the mergers are simulations 'no-winds' and
'BH' as described in \citet{nar07b}. The first merger snapshots
considered are when the molecular disks of the progenitors are
overlapped as the galaxies approach final coalescence.  We utilize
multiple snapshots throughout the evolution of the merger simulations
beyond this point to represent observed merging pairs caught in
various stages of evolution (e.g. close pairs such as the Antennae to
more evolved mergers such as Arp 220 and NGC 6240). All merger
snapshots are during the period of highest star formation activity,
when the galaxy can be considered a LIRG/ULIRG (model IR luminosities
for these merger simulations can be found in \citet{cha07a}). Sample CO
images of the merger simulations employed here, as well as some of the
model disk galaxies can be found in \citet{nar07b}. 

It is an important point to note that no particular choice or
combination of models affects the presented results. While some of the
disk galaxies used may nominally have larger circular velocities or
gas fractions than galaxies observed in the local Universe, because
the results are general, not including them does not change our results
or interpretation. We include a wide range of model galaxies to
increase the number statistics and dynamic range of our
simulations. The generality of our results will be shown more
explicitly in \S~\ref{section:origins} and
\S~\ref{section:predictions}.

\subsection{Non-LTE Radiative Transfer}
\label{section:radtrans}
The propagation of a line through a medium with lower density than the
line's critical density requires a full non-LTE
treatment. Specifically, in this regime, both collisions and radiative
processes contribute to the excitation and de-excitation of
molecules. This formalism for molecular line transfer has long been
applied to interpretations of galaxy observations via large velocity
gradient codes \citep[e.g.][]{golkwa74}, though has only recently been
incorporated in full 3 dimensions in galaxy-wide models
\citep{gresom06,nar06a,nar07a,nar07b,wadtom05,yam07}. We employ the
radiative transfer methodology of \citet{nar07b}, and refer the reader
to that work for full details. Here, we briefly summarize.

The radiative transfer is performed in two phases. First, the
molecular level populations (and consequently source functions) are
explicitly calculated using the 3D non-LTE code, {\it Turtlebeach}
\citep{nar06a,nar07b}. In this, a solution grid to the level
populations is guessed at, and model photons are emitted isotropically
in a Monte Carlo manner. The level populations are updated by assuming
statistical equilibrium, and balancing radiative and collisional
excitations, de-excitation and stimulated emission \citep{ber79}.  A
new series of model photons is emitted and the process is repeated
until the level populations are converged.

Once the level populations and source functions are known, the model
intensity can be found by integrating the equation of radiative
transfer along various lines of sight. Formally:
\begin{equation}
  I_\nu = \sum_{r_0}^{r}S_\nu (r) \left [ 1-e^{-\tau_\nu(r)} \right
  ]e^{-\tau_\nu(\rm tot)}
\end{equation}
where $I_\nu$ is the frequency-dependent intensity, $S_\nu$ is the
source function, $r$ is the physical depth along the line of sight,
and $\tau$ is the optical depth. 

The SPH outputs are smoothed onto a grid with $\sim$160 pc spatial
resolution for the radiative transfer.  In order to more accurately
describe the strongly density-dependent collisional excitation and
de-excitation rates, we model the gas in grid cells to be bound in a
mass spectrum of GMCs constrained by observations of Milky Way clouds
\citep{bli06}. The GMCs are modeled as spheres with power-law density
gradients, and radii given by the Galactic GMC mass-radius relation
\citep[e.g.][]{ros05,ros07,sol87}. These GMCs are placed randomly in
the grid cells (which we refer to as ``cells'' hereafter). Of the SPH
simulations utilized here in particular, the maximum volume-averaged
\htwo \ density seen in a cell was of order $\sim$1000 \cmthree,
though higher maximum densities are of course realized in the subgrid
GMCs, dependent on the mass of the GMC \citep[typically central core
densities reached $\sim$10$^6$-10$^7$ \cmthree \ in the nuclear
regions of star forming galaxies in our
simulations;][]{nar07b}. Because the sub-grid GMCs are posited on the
grid in post-processing, the SFR (Equation~\ref{eq:sfr1}) calculated
during the hydrodynamic simulations utilizes the volume-averaged
densities achieved at the hydrodynamic resolution (which is of order
$\sim$ 100 pc). More details concerning the subgrid formalism for
including GMCs on a galaxy-wide scale, and resolution tests may be
found in \citep{nar07b}.

Observational evidence suggests a range of power-law indices for GMCs,
ranging from $n$=1-2 \citep{and96,ful92,wal90}. Tests utilizing a
number of cloud density power-law indices within this range (in which
the total mass of the cell was conserved, and thus the central density
in the clouds allowed to vary) show that the results in this work are
not sensitive to this parameter choice. Similarly, in the study of
\citet{bli06} and \citet{ros07}, GMC mass spectrum indices were found
to range from $\gamma \approx$-1.4 to -2.8. Again, tests of mass
spectrum indices showed that the results of this paper are not
sensitive to parameter choices within the range of observational
constraints. We nominally employ indices of $n$=1.5 and $\gamma$=-1.8
for the cloud density and GMC mass spectrum power-laws \citep{ros07}.
This methodology allows us to faithfully capture the emission
processes from both dense molecular cores and more diffuse GMC
envelopes \citep{nar07b}.

We have benchmarked our radiative transfer codes against published
non-LTE radiative transfer tests \citep{van02}, and present the
results for these tests in \citet{nar06b}. Our methodology for
applying 3D non-LTE molecular line radiative transfer has shown
success in reproducing characteristic observed CO line widths,
morphologies, excitation conditions and intensities in isolated disk
galaxies, local ULIRGs and quasars from \zsim2-6
\citep{nar06a,nar07a,nar07b}.

The molecular gas mass fraction is assumed to be half, as motivated by
local volume surveys \citep[e.g.][]{ker03}, and the molecular
abundances set uniformly at Galactic values \citep{lee96}.  In this
work, typically $\sim$1$\times$10$^7$ model photons were emitted in
the non-LTE calculations per iteration, and we considered transitions
across 11 molecular levels at a time. The mass spectrum of GMCs had a
lower mass cutoff of 1$\times$10$^{4}\msun$ and an upper mass cutoff
of 1$\times$10$^6 \msun$.. The collisional rate coefficients were taken
from the {\it Leiden Atomic and Molecular Database} \citep{sch05}.

\section{Origin of Observed SFR-CO and SFR-HCN Slopes}
\label{section:origins}

\subsection{General Argument}

\begin{figure*}
\centerline{
\includegraphics[width=5.25cm]{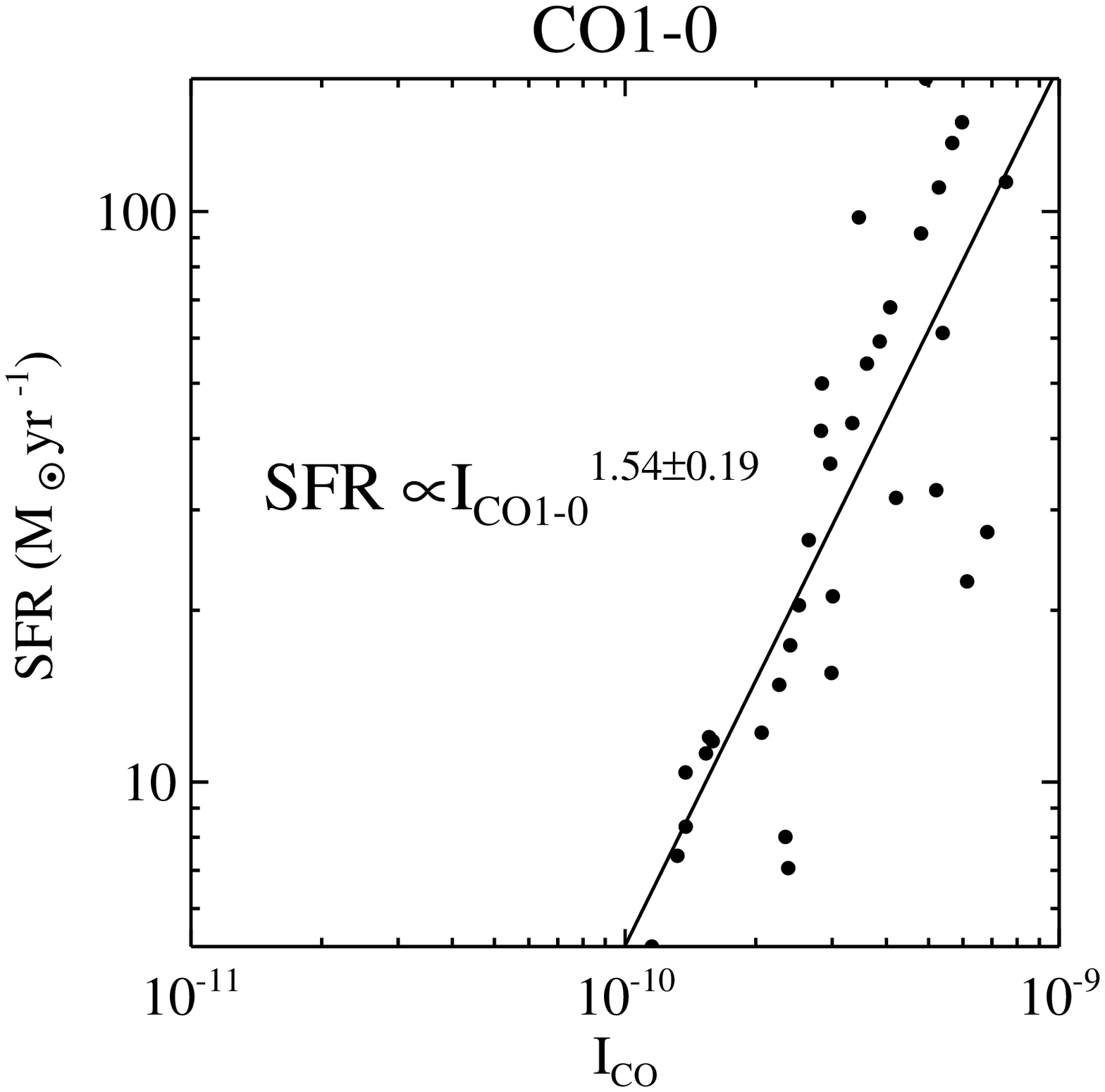}
\includegraphics[width=5.25cm]{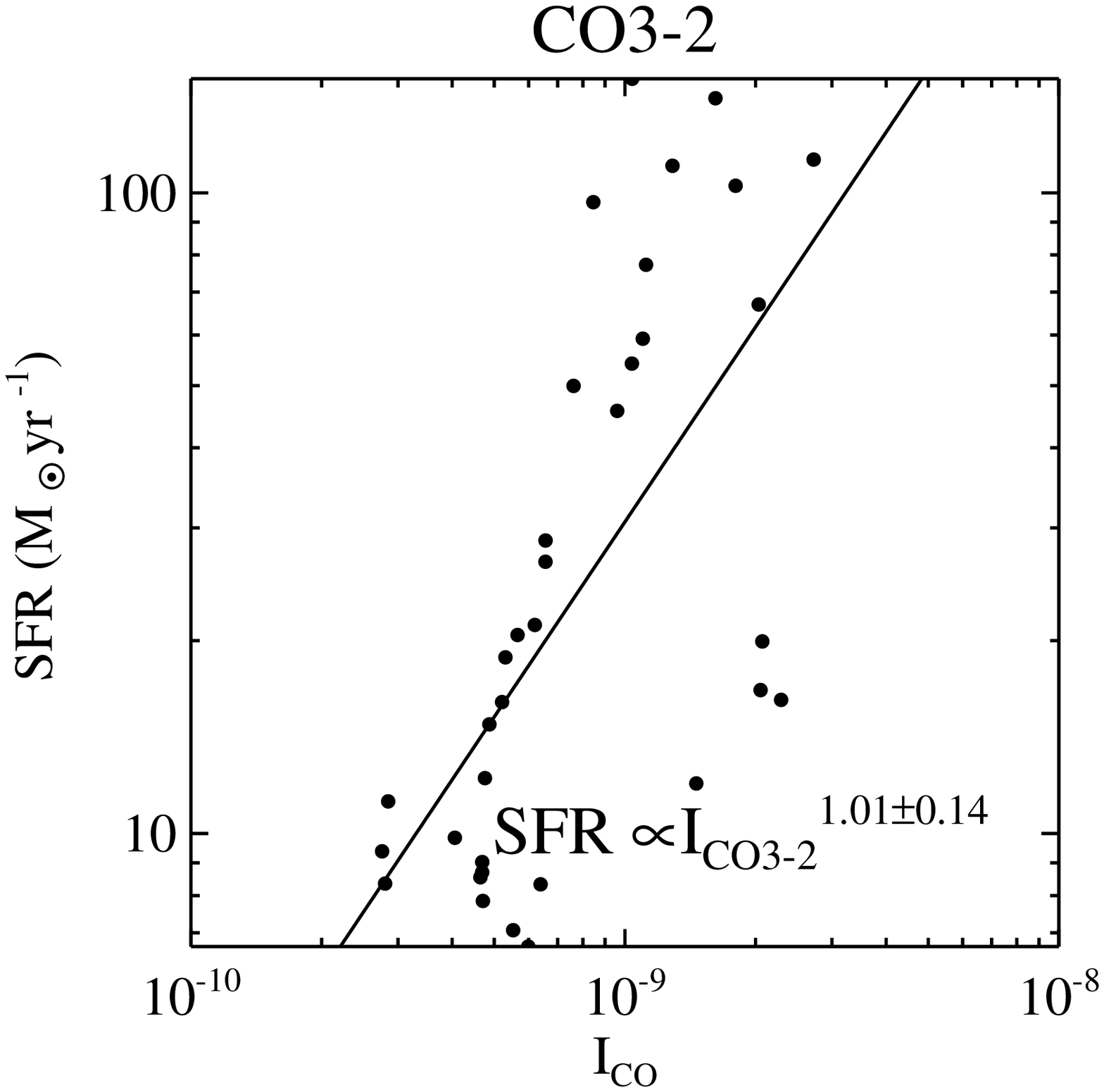}
\includegraphics[width=5.25cm]{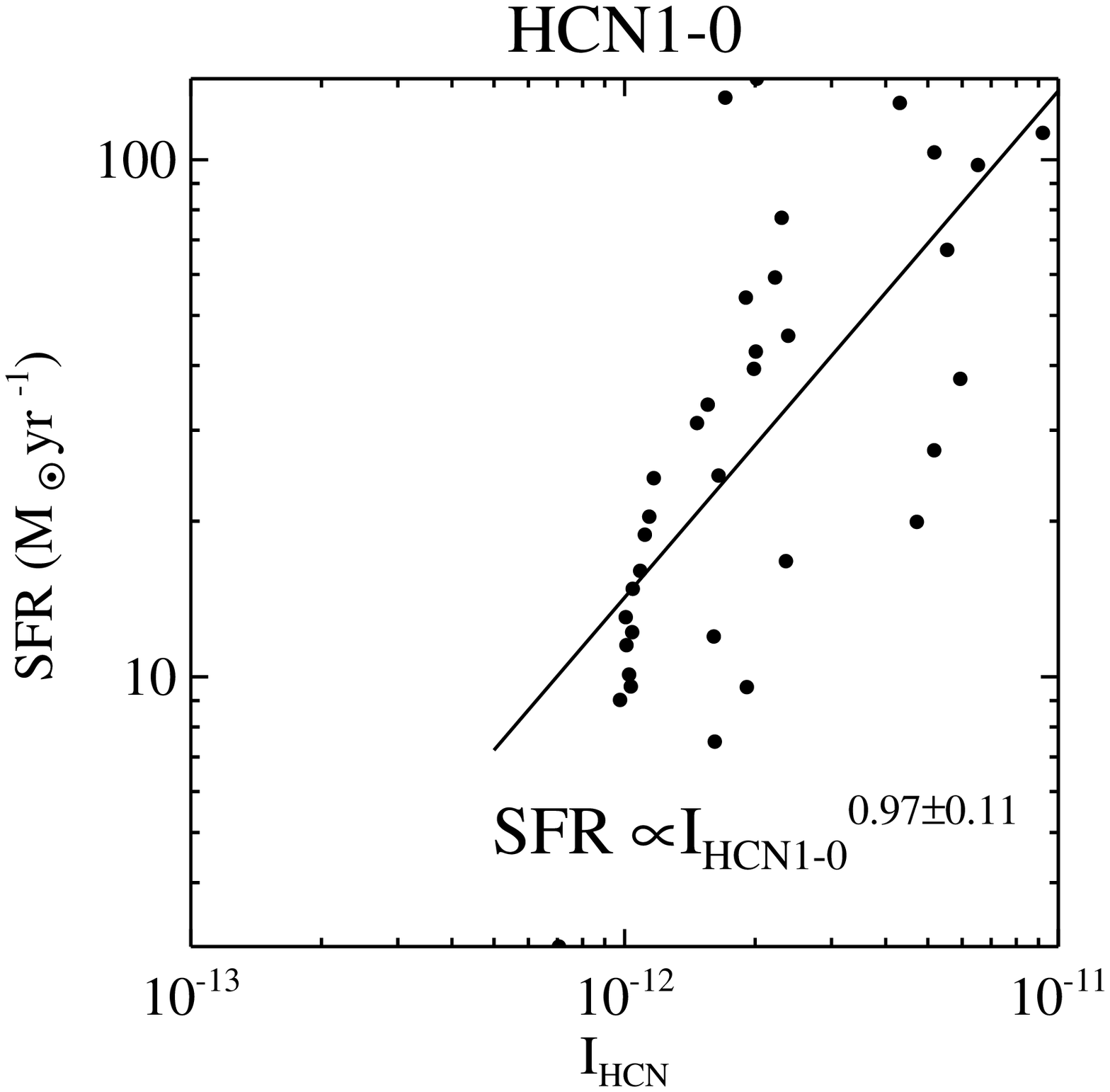}
}
\caption{Model results showing the relationship between SFR and CO
(J=1-0), CO (J=3-2) and HCN (J=1-0) emission in simulated galaxies.
The molecular line emission is derived for a randomly drawn set of
$\sim$35 galaxy snapshots of our model sample of $\sim$100, and is a
velocity-integrated line intensity. The randomly drawn set is
different for each plotted transition in order to illustrate the
generality of our results. The solid lines are the least-squares fit
to the plotted model results, and the fit is quoted in each panel.  We
randomly drew the sample of 35 galaxies 100 times, and quote the
1$\sigma$ dispersion in the derived slopes as the 'error' values. The
modeled relationships between SFR and molecular line luminosity are
consistent with the observations of \citet{gao04a,gao04b} and
\citet{nar05}. This figure takes the emission from the simulated
galaxies as a whole, thus simulating unresolved detections of the
galaxies. Thus it serves as a direct comparison to the observations by
\citet{gao04a,gao04b} and \citet{nar05}. While this particular plot
shows simulated unresolved observations of galaxies, we refrain from
using traditional molecular line ``luminosity'' units as future
discussion will center largely around the physical intensity
associated with individual emitting gas cells in the simulations. The
units of intensity are in erg s$^{-1}$ cm$^{-2}$
Hz$^{-1}$. \label{figure:sfrco}}
\end{figure*}

In Figure~\ref{figure:sfrco} we plot the SFR-CO relations and SFR-HCN
(J=1-0) relation as derived from our models of isolated disk galaxies
and mergers with best fitting slopes overlaid. Here, we show the
emission from a sample of model galaxies as a whole (e.g.  simulated
unresolved observations of galaxies). We plot the line intensities,
though note that they are proportional to molecular line luminosity
($L'$, \citet{solvan05}), the standard quantity reported in the
observational literature\footnote{$L'\approx \pi/$(4 ln 2)$\theta_{\rm
mb}^2 I_{\rm mol} d_L^2$(1+\z)$^{-3}$ where $\theta_{\rm mb}$ is the
main beam angular size, $I_{\rm mol}$ is the molecular line intensity,
$d_L$ is the luminosity distance, and \z \ is the redshift
\citep{gao04a}.}. Included in the plot are a random sampling of 35 of
the 15 disk galaxies and $\sim$80 merger snapshots in our parent
sample.  The points are all different simulated galaxies, and not
simply multiple viewing angles of the same galaxy. We first note the
general agreement of the best fit slopes and dispersion with observed
relations by \citet{gao04a,gao04b}, \citet{nar05}, and
\citet{san91}. The dispersion in \lmol \ at a fixed SFR arises owing
to variance in the mean density of the emitting gas in galaxies at a
given SFR.  We will discuss the physical reasoning behind the slopes
shortly.  Second, we note that we have utilized a randomly drawn
sample of our simulated galaxies to demonstrate the generality of our
results. From this, we see that any given combination of model disk
galaxies and/or merger snapshots give roughly the same fits. We have
quoted as ``error'' values the 1$\sigma$ dispersion in 100 random
draws of 35 model snapshots.  The generality of these results occurs
because the origin of the SFR-molecular line relations arise from the
nature of the molecular emission from the individual galaxies
themselves.

To see this, consider a galaxy which is forming stars at rate:
\begin{equation}
\label{eq:sfrmh2}
SFR \propto \rho^N 
\end{equation}
where $\rho$ is the mean molecular gas mass density.  The relationship
between the SFR and the luminosity of a molecular line from a cell of
clouds
\begin{equation}
\label{eq:sfrlmol}
SFR \propto L_{\rm molecule}^\alpha
\end{equation}
is dependent on the relationship between the molecular line luminosity
and the mean gas density in a cell of clouds:
\begin{equation}
\label{eq:lmolmh2}
L_{\rm molecule} \propto \rho^{\  \beta}
\end{equation}
where $\beta$=N/$\alpha$. Therefore, for a given Schmidt law index,
$N$, the root issue in determining how the SFR relates to molecular
line luminosity is understanding how the molecular line traces
molecular gas of different densities ($\beta$). 


The global relationship between molecular line luminosity and mean gas
density traced ($\beta$) is driven by how the critical density of the
molecular line compares with the density of the bulk of the clouds
across the galaxy. In short, lines which have critical density below
the mean density of most of the emitting gas cells will be thermalized
and rise linearly with increasing cloud density. These lines will
consequently have SFR-line luminosity relations ($\alpha$) similar to
the Schmidt index controlling the SFR.

\begin{figure}
\scalebox{0.8}{\rotatebox{90}{\plotone{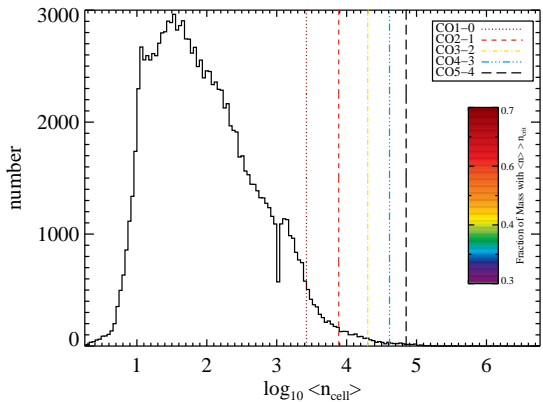}}}
\caption{Distribution of mean cloud densities throughout our fiducial
disk galaxy with circular velocity $v$=160 \kms and gas fraction
$f$=0.4. The lines overlaid denote the location of critical densities
of common CO transitions, and the color of each line indicates the
fraction of gas mass with density above each critical density (with
color scale on right). While most of the gas mass is above the
critical density of the lowest lying transition, CO (J=1-0), higher
lying transitions (e.g. CO J=3-2) may be subthermally populated
through much of the galaxy.\label{figure:meandens_ncrit}}
\end{figure}

Lines which have critical density well above the mean density of most
of the emitting gas cells will be thermalized in only a small fraction
of the gas. To see this, in Figure~\ref{figure:meandens_ncrit}, we
plot the distribution of mean cell densities for a fiducial disk
galaxy (with 40\% gas fraction and 160 \kms circular velocity; this
will be a fiducial disk galaxy which we center the remainder of the
discussion around as a reference point, though the results are
general)\footnote{In order to account for the destruction of molecules
in photodissociation regions \citep[e.g.][]{hol99}, we do not consider
emission from regions with column density $\la$1.5$\times$10$^{21}$
\cmtwo. This typically corresponds to cells with mean cloud density
$\sim$50 \cmthree \ in our simulations.}. We additionally plot the
critical densities of a sample of CO lines overlaid as an indicator of
the relative gas fractions above given critical densities. In the case
of high critical density tracers such as e.g. CO (J=3-2), a small
fraction of the gas is thermalized. As the mean density increases, the
fraction of thermalized gas (and consequently photon production)
increases superlinearly, driving an e.g. SFR-CO (J=3-2) relation with
index less than that of the Schmidt index. This is similar to the
physical mechanism driving the SFR-\lmol \ relation in star forming
GMCs by \citet{krumthom07}.

An important question to understand is, how does this global model for
the observed SFR-\lmol \ relations \citep[also considered by
][]{krumthom07} manifest itself when considering the actual transfer
of molecular line photons in physical models of isolated galaxies and
galaxy mergers?  In particular, high critical density photons from the
dense, thermalized nucleus may be reabsorbed by diffuse gas along the
line of sight. How, then, considering this redistribution, is the
observed SFR-\lmol \ relation realized? In short, in galaxies where
$\bar{n} \ll n_{\rm crit}$, large amounts of diffuse gas along the line
of sight are subthermally (radiatively) excited owing to line
trapping. Emission from subthermally excited gas can be thought of as
a scattering process in that the high critical density photon
originates in dense, thermalized molecular gas, and is redistributed
to more diffuse gas.  Subthermal emission from radiatively excited
cells tends to have a superlinear \lmol-$\bar{n}$ relation as the
relatively diffuse gas owes its luminosity to molecules who have been
pumped by radiation from denser regions. As such, the radiation is
indicative of denser, thermalized regions where the photons were
originally created, but the diffuse gas along the line of sight
reduces the sightline-averaged mean density. In these cases, the
SFR-line luminosity relation ($\alpha$) has an index less than that of
the underlying Schmidt law for the galaxy.  There are details, of
course, specific to the relationship between CO (J=1-0), CO (J=3-2)
and HCN (J=1-0) luminosities and gas density traced which require a
more in depth analysis, but these underlying themes are robust. This
is inherently an alternative but equivalent way of viewing the
previous point. The net number of high critical density photons
depends on the quantity of dense, thermalized gas. The bulk of the gas
along the line of sight (which is responsible for the line trapping)
is diffuse, resulting in a superlinear \lmol-$\bar{n}$ relation along
the line of sight.

We devote the remainder of this section to examining the driving
mechanisms behind the relationship between line luminosity and mean
gas density ($\beta$) in greater detail, and utilize these to
formulate a general model for observed SFR-molecular line luminosity
relations in galaxies. We will utilize the aforementioned fiducial
disk galaxy as a reference point. We will couch the discussion (at
least for the high critical density tracers - e.g. HCN J=1-0 and CO
J=3-2) in terms of both the global relationship between $\bar{n}$ and
$n_{\rm crit}$, as well as the redistribution of photons from
thermalized gas to diffuse, subthermally excited cells. The reason we
do this is the following.  The SFR-\lmol \ relation in galaxies is
{\it globally} set by the number of high critical density photons
produced as a function of the mean density of the emitting clouds
along the line of sight (e.g. the relationship between $\bar{n}$ and
$n_{\rm crit}$; Figure~\ref{figure:meandens_ncrit}). While these high
critical density photons are originally produced in the small number
of thermalized cores in galaxies, this does not necessarily reflect
the final emitting surface in the galaxy. The photons are
redistributed from dense cores to being absorbed and re-emitted by
diffuse gas along the line of sight.  The details of how the SFR-\lmol
\ relation is driven in a realistic model for a galaxy are subtle, and
require more analysis than the global arguments presented
thusfar. Moreover, the large quantities of subthermally excited gas is
a {\it direct observational consequence} of this physical model for
the observed SFR-line luminosity relations. For these reasons, we
continue forward framing the understanding of the observed relations
between SFR and line luminosity in terms of both global properties as
well as the subthermally excited gas along the line of sight. That
said, the details outlined in the remainder of this section are just
that. The remainder of the predictions presented in this paper are not
dependent on the finer points regarding the excitation patterns in
galaxies.

\subsection{Detailed Understanding of an Individual Galaxy}

We first note that a significant portion of the intrinsic luminosity
in the galaxy escapes. While the optical depths can become large
locally, velocity gradients across the galaxy serve to shift native
absorption profiles out of resonance with the emission line on larger
scales.  Specifically, we see $\sim$60\% of the emission across most
CO and HCN transitions escape the galaxy. The gas cells contributing
to emission that escapes are of a fixed size, and distributed across
the galaxy.  These cells exhibit a wide range of mean cloud
densities. We therefore focus on the properties of individual gas
cells in our simulations (of all masses and densities) in order to
build an understanding of how their summed properties drive the global
line luminosities-gas density relation, and the consequent
SFR-molecular line luminosity relation.

 The observed SFR-molecular line relations \citep{gao04a,gao04b,nar05}
are representative of three regimes: one in which the mean density of
the bulk of the emitting clouds is higher than the line's critical
density; one in which the mean density is lower than the line's
critical density; and one in which the mean density is lower than the
line's critical density and the line considered is a ground-state
transition. We explore these three cases here with respect to our
fiducial disk galaxy (though note again that the results are
general).

\begin{figure*}
\centerline{
\includegraphics[width=5.25cm]{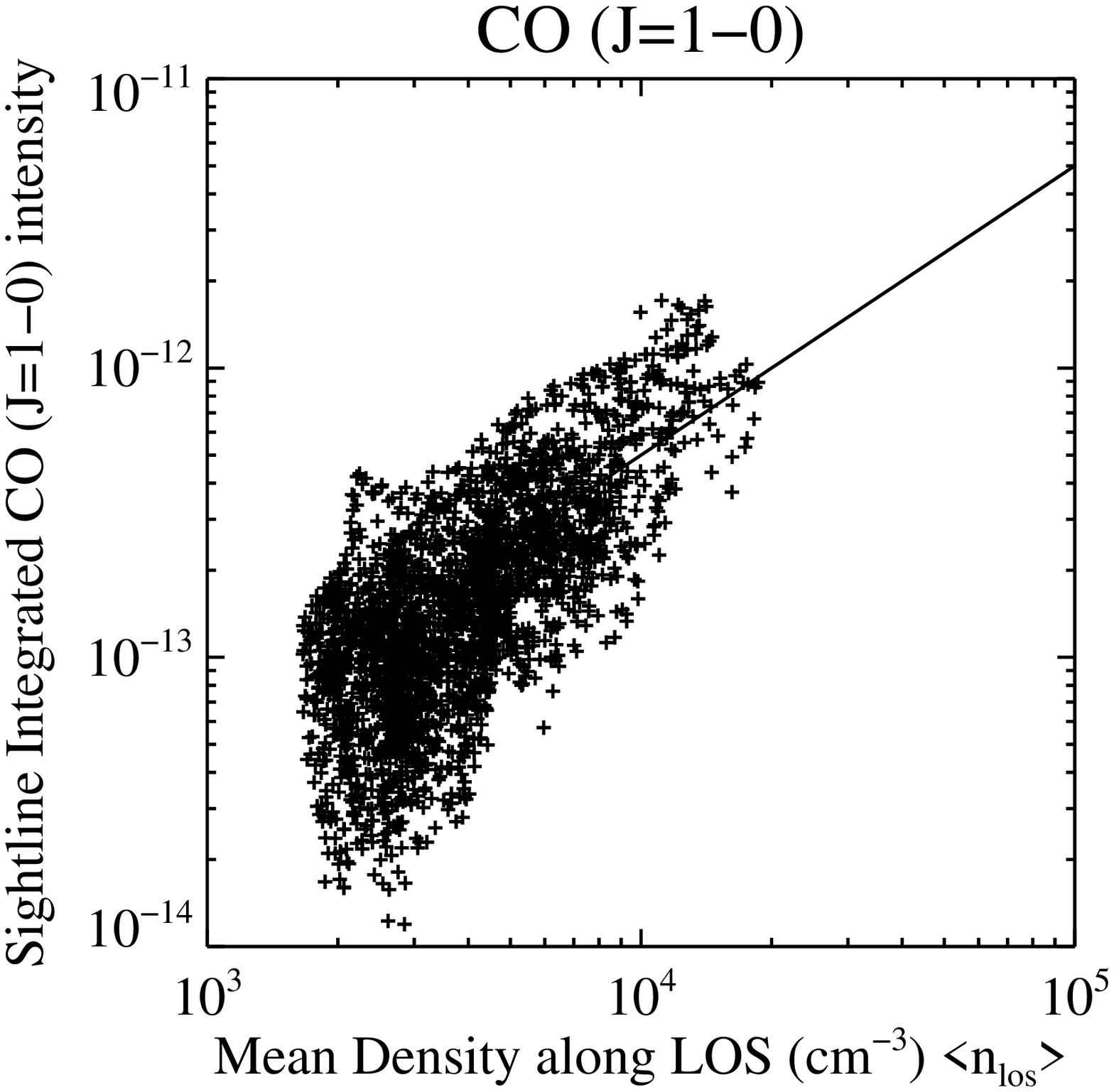}
\includegraphics[width=5.25cm]{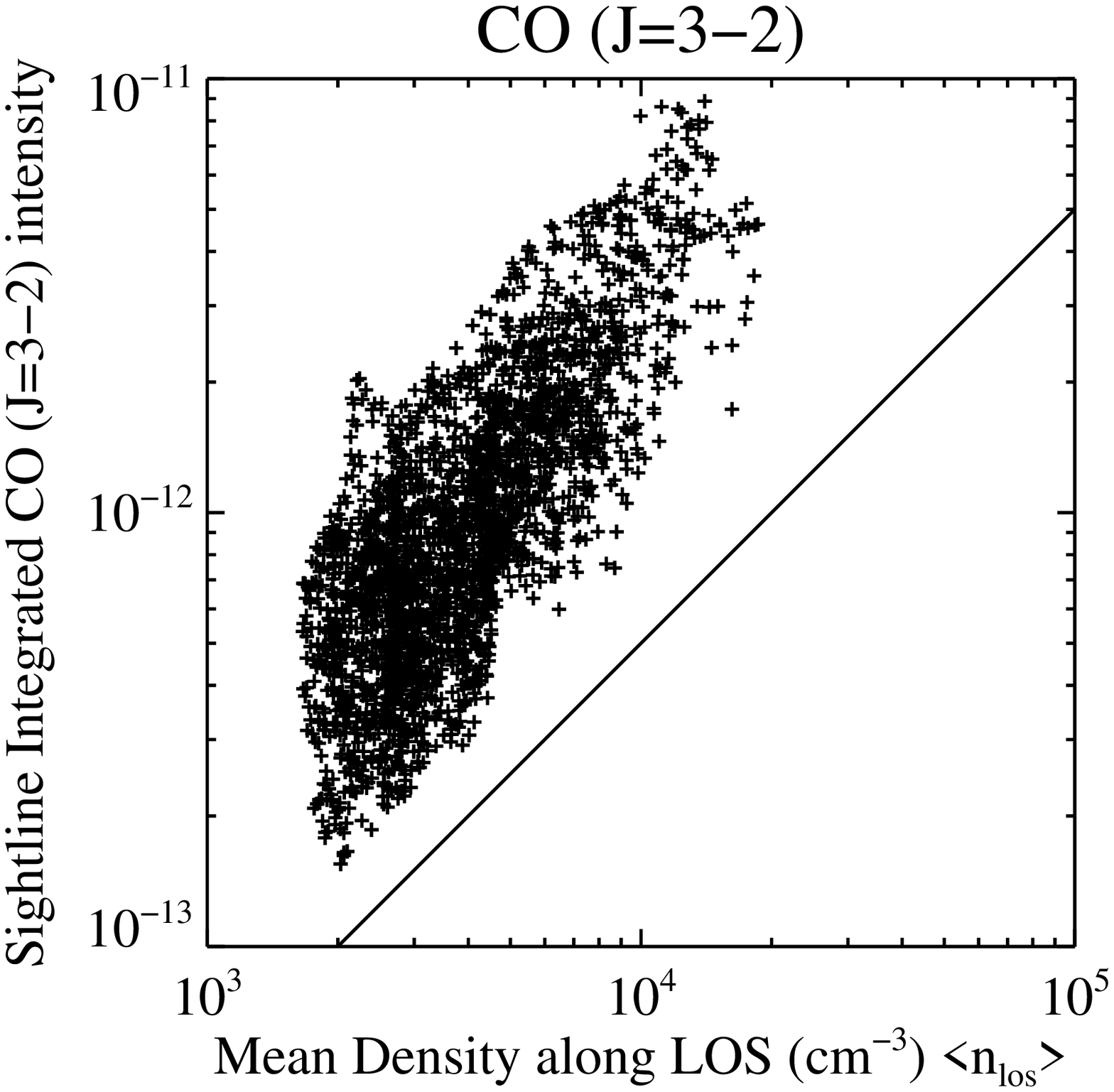}
\includegraphics[width=5.25cm]{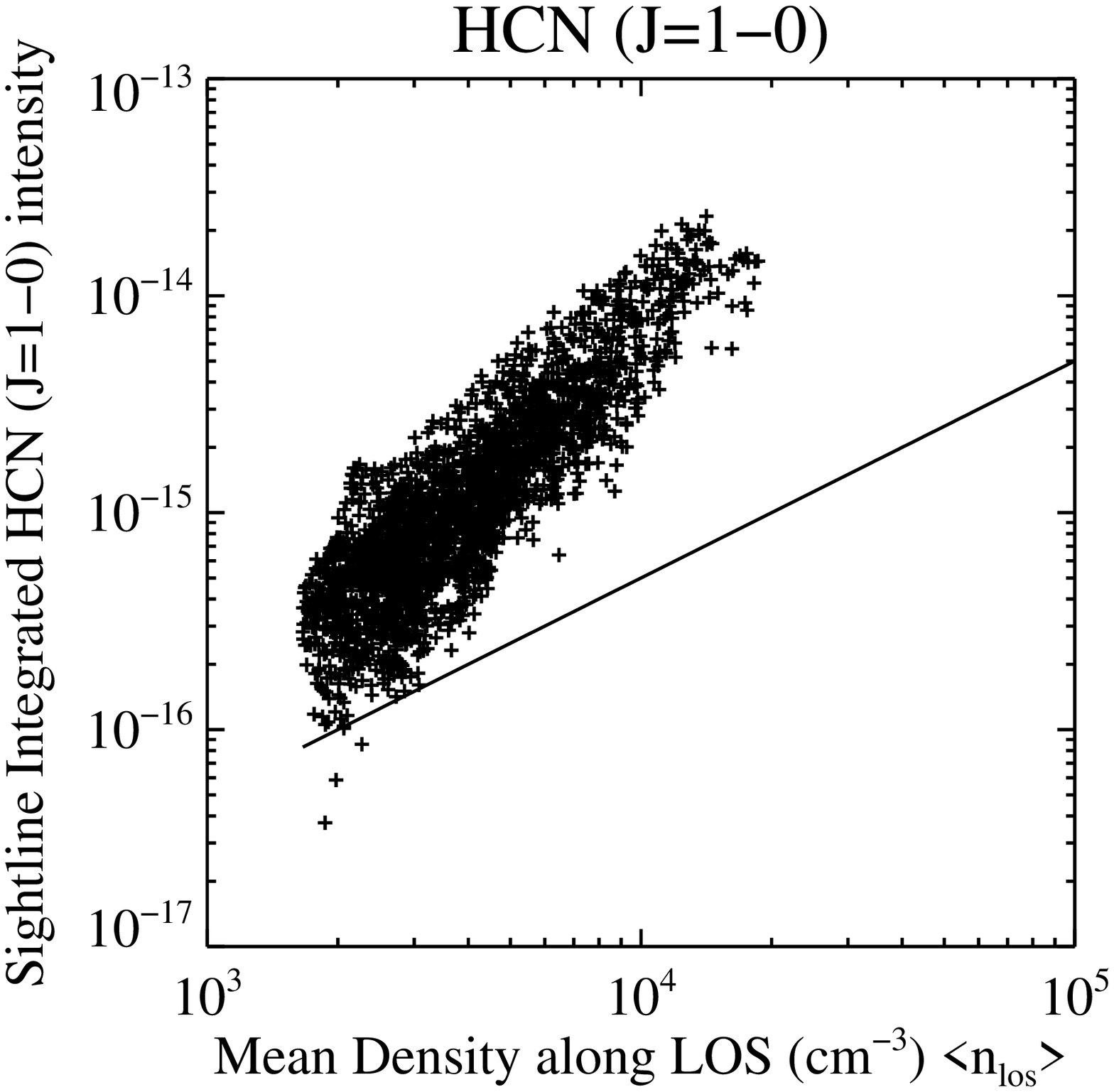}
}
\caption{Velocity integrated molecular line intensities (erg s$^{-1}$
cm$^{-2}$ Hz$^{-1}$ - \kmsend) along 50$^2$ sightlines for a fiducial
disk galaxy versus mean density along the line of sight.  The crosses
represent {\it individual sightlines} (with implied spatial resolution
of $\sim$160 pc) for an individual simulated disk galaxy. The quoted
mean density is the average along each sightline, and the
velocity-integrated intensity is the emission that escapes along each
sightline. The disk galaxy has a circular velocity of 160 \kmsend, and
a gas fraction of 40\% (though the results are general). The solid
lines show linearity, and the normalization is arbitrary. The CO
(J=1-0) emission traces the mean density along the line of sight
roughly linearly, while the higher critical density tracers trace mean
density superlinearly. These results are expected in order to recover
the observed SFR-line luminosity relations
(Equations~\ref{eq:sfrmh2}-\ref{eq:lmolmh2}).\label{figure:lvmeandens_los}}
\end{figure*}

\begin{figure*}
\centerline{
\includegraphics[width=6.2cm]{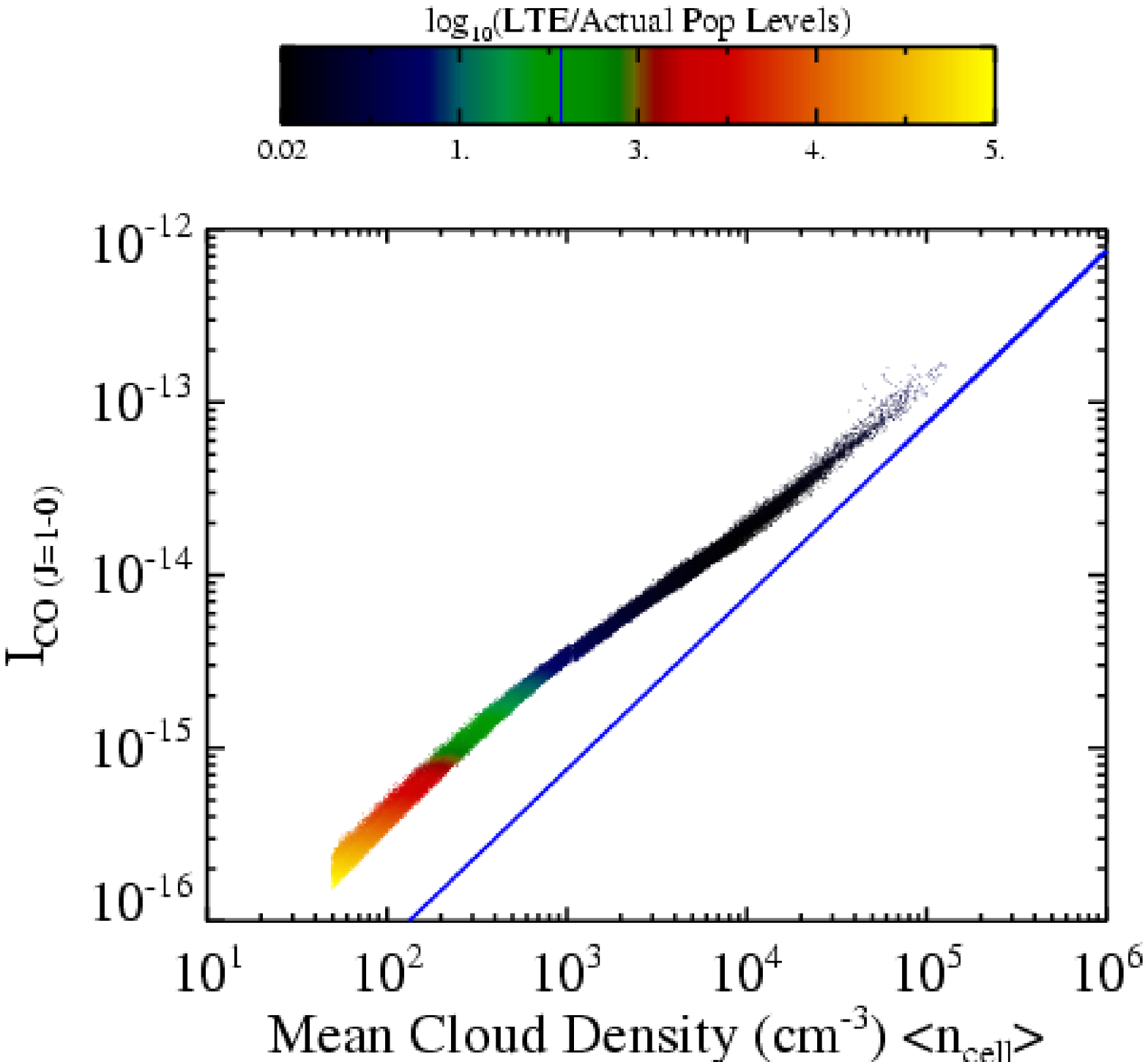}
\includegraphics[width=6.2cm]{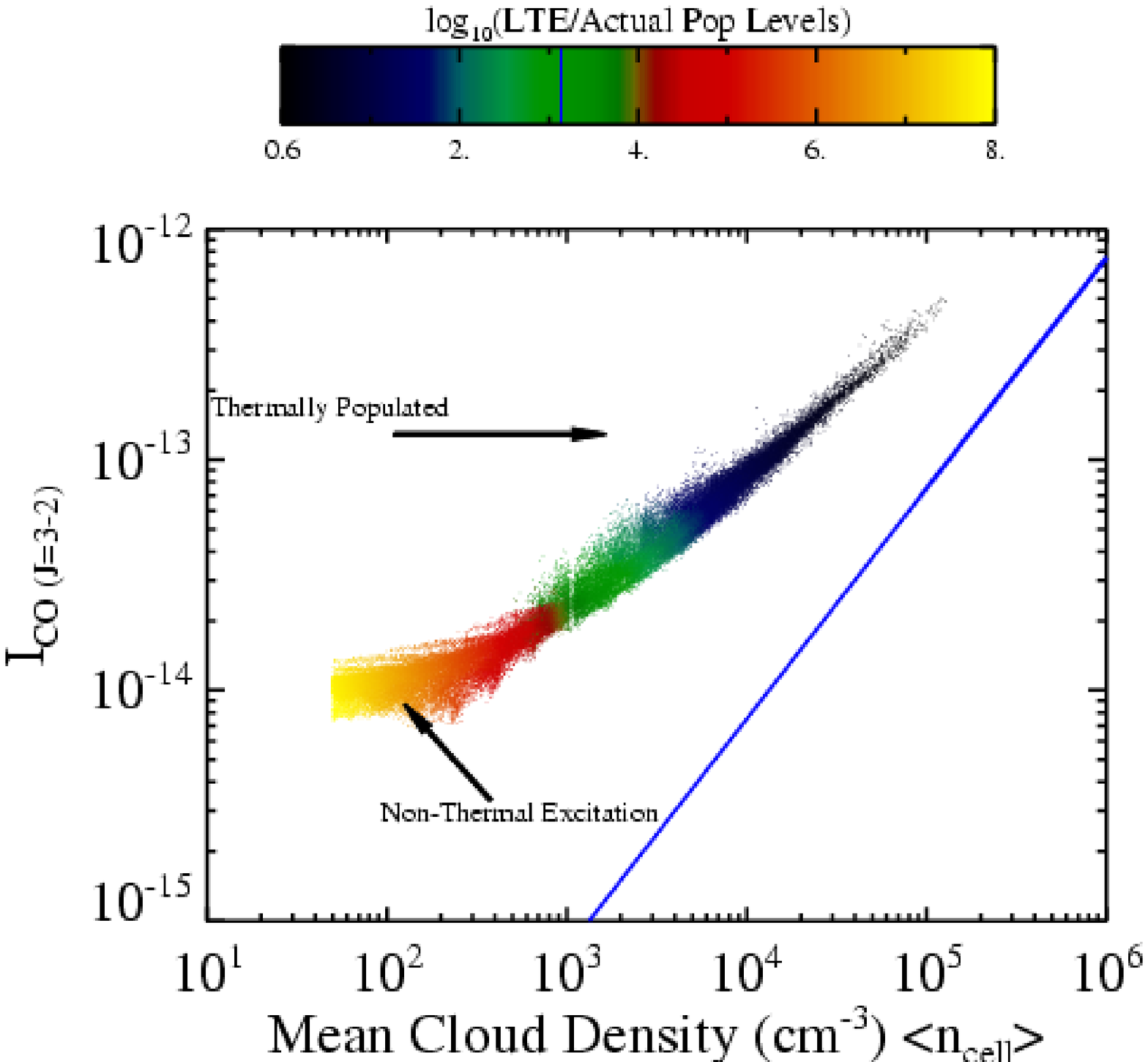}
\includegraphics[width=6.2cm]{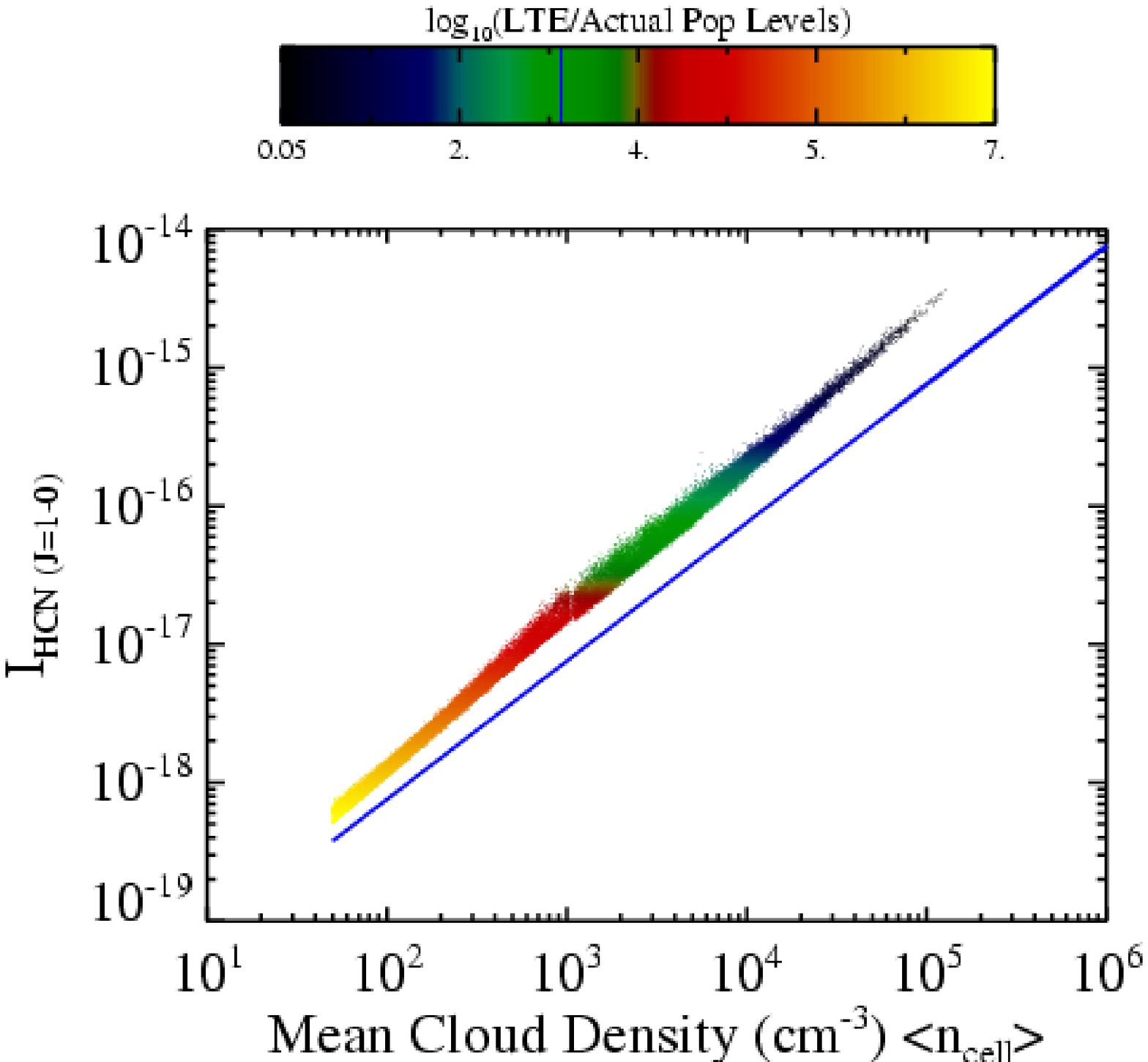}
}
\caption{CO (J=1-0), (J=3-2) and HCN (J=1-0) intensity versus mean
cloud density in our fiducial disk galaxy on a cell by cell
basis. Each cell typically contains a complex of GMCs.  The quoted
mean density is the mean density of the 50$^3$ cells in our sample
grid. The blue line shows linearity, and the colors in the points show
the ratio of LTE level populations to actual level populations for the
upper J level of each transition. See text for details on how this
drives the observed SFR-line luminosity relations.  The units of
intensity are in erg s$^{-1}$ cm$^{-2}$ Hz$^{-1}$ which are
proportional to the standard observed $L'$ molecular line luminosity
units.
\label{figure:lcovmeandens}}
\end{figure*}

In Figure~\ref{figure:lvmeandens_los}, we show empirically how the
line of sight velocity-integrated intensity from CO (J=1-0), CO
(J=3-2) and HCN (J=1-0) relate to the sightline-averaged mean density
along 50$^2$ sightlines for the fiducial disk galaxy\footnote{Note
that this is now plotting the \lmol-$\bar{n}$ relation for {\it
individual sightlines} for a single disk galaxy (with implied spatial
resolution equal to that of the simulation of $\sim$160 pc). This is
in contrast to Figure~\ref{figure:sfrco} which plotted simulated
unresolved observations of galaxies as a whole.}. The solid line in
each case shows linearity, with arbitrary normalization. In
Figure~\ref{figure:lcovmeandens}, we plot the line luminosity versus
mean density on a cell by cell basis for the CO (J=1-0), (J=3-2) and
HCN (J=1-0) transitions from our fiducial disk galaxy. We remind the
reader that a 'cell' is a grid cell in our fiducial galaxy which
contains potentially numerous GMCs whose effects are simulated in a
subgrid manner \citep{nar07b}. We will refer to these two figures
throughout the forthcoming discussion.

\subsubsection{$\bf n_{\rm crit} \ll \bar{n}$} 
\label{section:co10}
We first consider the regime in which the mean density of the bulk of
the emitting clouds is typically higher than the line's critical
density. We see empirically from the left panel of
Figure~\ref{figure:lvmeandens_los} that the CO (J=1-0) luminosity
traces sightlines of increasing mean density linearly in our fiducial
galaxy ($\beta = 1$). Utilizing
equations~\ref{eq:sfrmh2}-\ref{eq:lmolmh2}, this translates into a
globally observed relation in a sample of galaxies in which the SFR
relates to \lmol \ with an index similar to the underlying Schmidt
index (here, $N$=1.5). What causes this linear relation between \lmol
\ and $\bar{n}$?

At the critical density, the line is approaching thermalization and,
 in the limit that constant molecular abundances apply (with respect
 to \htwo), the clouds are typically optically thick. In this
 scenario, the emission arises primarily from the outer (lower column
 density) regions of clouds, and serves as a measure of the number of
 emitting clouds in a given cell. While groups of clouds are locally
 optically thick, on a galaxy-wide scale they are optically
 thin. Thus, the velocity dispersion of the bulk of clouds along a
 given line of sight has the effect of spreading the molecular line
 emission out in frequency space, permitting physically overlapping
 clouds to be counted. Because the mean cloud density increases with
 number of clouds in a given cell, in the optically thick limit, a
 roughly linear relation between molecular line emission and mean
 cloud density is natural.

In Figure~\ref{figure:lcovmeandens} (left panel), we illustrate this
effect by showing the CO (J=1-0) intensity as a function of mean cell
density in our fiducial galaxy.  The points are individual grid cells
containing (potentially) numerous GMCs. Overlaid in color is the level
of thermalization.  The bulk of the CO (J=1-0) emission comes from
cells that are either in LTE or nearly thermalized.  We therefore
arrive at the conclusion that transitions that have critical density
well below the mean critical density of most emitting clouds will be
roughly thermalized, and the SFR-line index will be similar to the
underlying Schmidt-law index. When an underlying Schmidt index of
$N$=1.5 is in place, the SFR-\lmol \ relation matches observations
well. This is similar to the explanation posited by \citet{krumthom07}
for their models of GMCs.

).

\subsubsection{$\bf n_{\rm crit} \gg \bar{n}$}

\begin{figure}
\plotone{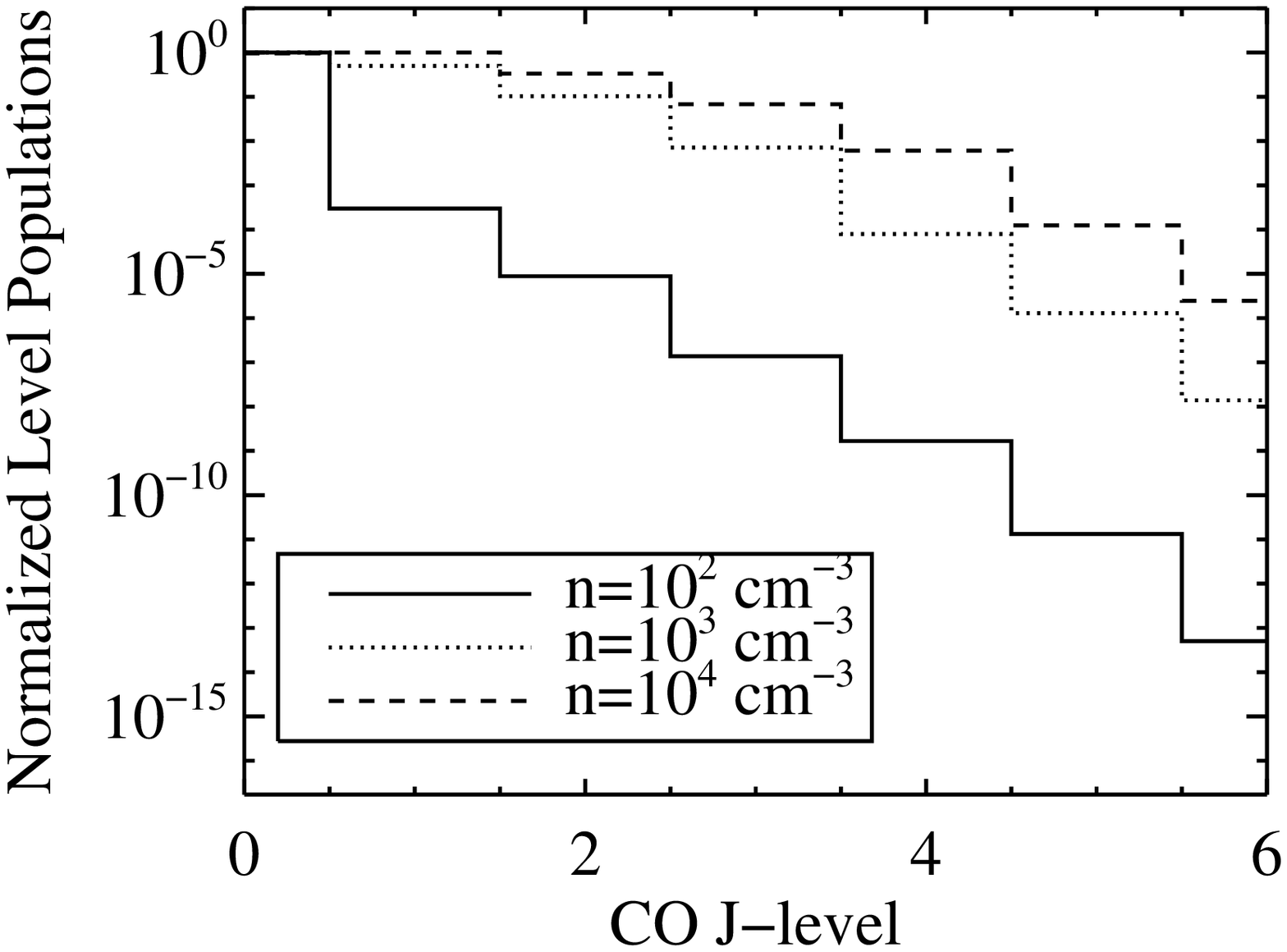}
\caption{ Normalized level population distributions for three gas
cells of varying mean density in our sample disk galaxy.  In the
lowest density cell ($\bar{n}$ $\sim$10$^{2}$\cmthree), all J$\geq$1
states are subthermally populated, and rising with increasing
density. The J=1 level is increasing with respect to roughly fixed J=0
populations with rising cloud density, resulting in increasing CO
(J=1-0) flux with rising mean cloud density. Conversely, in the
lower density curves, while the level populations from higher lying
levels are also increasing with mean cloud density, the are not
increasing {\it with respect} to each other (e.g. the relatively
constant J=3/J=2 level population ratio from the $\bar{n}$
$\sim$10$^{2}$\cmthree \ cell through the $\bar{n}$ $\sim$10$^{3}$\cmthree
\ cell). Here, the flux from e.g. the J=3-2 line will remain constant
with increasing mean cloud density. This occurs until the J=2 level
approaches thermalization and the J=3 populations begin to increase
with respect to the J=2 populations.
\vspace{0.5cm}
\label{figure:poplevels}}
\end{figure}

If the critical density of the transition is toward the high density
tail of the gas density distribution function, as one considers
galaxies of increasing mean density, the luminosity from that
molecular transition will increase superlinearly. This owes to an
increase in thermalized gas both from an increased quantity of gas in
the galaxy, as well as an increased fraction of gas above the critical
density.  This effect is seen in simulated global unresolved
observations of galaxies (which show effectively a single, unresolved
sightline in a galaxy) as well as simulated higher resolution
observations. We illustrate this intuitive argument in the middle and
right panels of Figure~\ref{figure:lvmeandens_los} where we show
50$^2$ observed sightlines through our fiducial disk galaxy.

The superlinear \lmol-$\bar{n}$ relation in high critical density tracers
drives a relation between SFR and \lmol \ which has index less than
the Schmidt index (Equations~\ref{eq:sfrmh2}-\ref{eq:lmolmh2}).  This
is similar to the interpretation advocated by \citet{krumthom07} in
their models of the SFR-\lmol \ relation in GMCs. 

As discussed before, this physical model drives the global SFR-\lmol \
relation in our model galaxies. That said, the connection between this
and how the \lmol-$\bar{n}$ relation is realized on a cell by cell
basis when considering physical models of galaxies, and the effects of
radiative transfer and light redistribution is more subtle. In the
remainder of this section, we examine the two case examples of
$\bar{n} \ll n_{\rm crit}$ of CO (J=3-2) and HCN (J=1-0) in detail.

\paragraph{Non Ground-State Transition}

We first consider the case in which the gas density is in large part
below the critical density of the emission line (which is not a ground
state transition - e.g. CO J=3-2) on a cell by cell basis for our
fiducial disk galaxy. In the lowest density cells, high critical
density tracers such as CO (J=3-2) are subthermally excited. Unlike
the case where $\bar{n} \gg n_{\rm crit}$, the emission from the lower
mean density cells does not drop monotonically with decreasing mean
cloud density. Rather, the molecular excitation (and consequent
emission) is supported by radiative excitation from neighboring cells
with higher mean density.

The cells with the lowest mean density show a relatively constant flux
level which is higher than would nominally be allowed if collisions
alone drove the molecular excitation
(Figure~\ref{figure:lcovmeandens}).  To see this, consider the
dependence of the cell's intensity on the molecular level populations:
\begin{equation}
\label{eq:sourcefunction}
I_{\nu}=\frac{n_{u}A_{ul}}{(n_{l}B_{lu}-n_{u}B_{ul})} \, ,
\end{equation}
where $n_u$ and $n_l$ are the upper and lower state level populations,
and $A_{\rm ul}$, $B_{\rm lu}$ and $B_{\rm ul}$ are the Einstein
coefficients for spontaneous emission, absorption, and stimulated
emission, respectively. In subthermally populated levels, $n_l  \gg
n_u$, and consequently $I_{\nu} \propto n_u/n_l$.  When both $l$ and
$u$ are subthermal, $n_l$ and $n_u$ increase in lock step
superlinearly with increasing mean cloud density. This owes to a
combination of the effects of collisional excitations (which increases
linearly with density) and the additional contribution of radiative
excitations. Because $n_l$ and $n_u$ both increase monotonically with
gas density, there will be little change in the intensity as a
function of increasing gas density. Consequently, the value of the
intensity is roughly constant at ($n_uA_{ul}$)/($n_lB_{lu}$) where
$n_u$ and $n_l$ are inflated (above the effects of collisional
excitation alone) by line trapping (Figure~\ref{figure:lcovmeandens}).

At higher mean cloud densities, when the collisions begin to
contribute significantly to the excitation processes for the $l$
state, $n_u$ continues to increase superlinearly with increasing
density whereas $n_l$ increases only linearly. Thus, $n_u$ increases
with respect to $n_l$ with increasing mean cloud density, and the line
intensity begins to rise with cloud density
(Equation~\ref{eq:sourcefunction}). Emission in these cells is coming
both from subthermally excited gas, and the small fraction of
thermalized gas (in cloud cores) in these relatively low mean density
cells. At the highest mean cloud densities, the $l$ and $u$ states are
both thermally populated, and the emission rises linearly with
increasing mean density as in \S~\ref{section:co10}. In
Figure~\ref{figure:poplevels}, we plot an illustrative example of how
the level populations evolve with increasing mean cloud density by
showing the CO level population distributions for three cells with
different mean densities in our fiducial disk galaxy. These rates of
level population increase with increasing mean gas density translates
to the relationship between $I_{\rm CO}$ and gas density for the CO
(J=3-2) transition as shown in Figure~\ref{figure:lcovmeandens}
(middle panel).

 At first glance, Figure~\ref{figure:lcovmeandens} suggests that the
CO (J=3-2) intensity traces the gas density ($\beta$) sublinearly,
which is contradictory to what we would expect given the nearly linear
SFR-CO (J=3-2) index ($\alpha$;
Equations~\ref{eq:sfrmh2}-\ref{eq:lmolmh2}). However, the important
quantity to consider is the total integrated intensity summed along
sightlines through the galaxy (Figure~\ref{figure:lvmeandens_los}).
 The gas in the low density subthermally populated regions along the
line of sight is excited by emission from warmer, higher density gas,
and thus has a characteristic intensity reflective of a higher
brightness temperature than the meager densities in these gas cells
would normally allow for via collisional excitation alone. Conversely,
emission from higher density gas is more representative of the dense
regions the photons originate from. If one could see directly into the
thermalized nucleus, the CO (J=3-2) emission would be characteristic
of the density of gas traced, and the consequent $I_{\rm CO}$-gas
density relation would be linear
(e.g. \S~\ref{section:co10}). However, line of sight observations of
the galaxy contain a contribution both from the dense nucleus, and the
subthermal cells along the LOS which have a rather high \lmol/$\bar{n}_{\rm
cell}$ ratio. The contribution to the emission from lower mean
density regimes along the line of sight results in a total superlinear
relation between intensity and line of sight mean gas density
($\beta$).

\begin{figure}
\scalebox{0.8}{\rotatebox{90}{\plotone{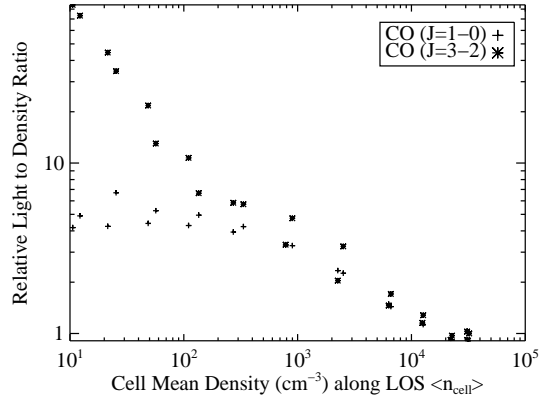}}}
\caption{Ratio of intensity to gas density in gas cells along a line
of sight that goes through the nucleus of the sample galaxy for the CO
(J=1-0) and CO (J=3-2) transitions.  Both the intensity and density
curves are normalized to their respective maximum values (which both
occur in the densest cell), so the expected light to density ratio is
unity at the maximum density. For the CO (J=1-0) transition, most of
the gas is thermalized along the line of sight, so the relative light
to density ratio is nearly unity for most clouds. For the CO (J=3-2)
transition, diffuse gas along the sightline contributes proportionally
more luminosity than thermalized cells owing to radiative excitations
by denser gas in the vicinity. When comparing to
Figure~\ref{figure:meandens_ncrit}, it is evident that these lower
density clouds are numerous along a given sightline. The summed
contribution of the intensity from these low density cells to the
total integrated intensity causes the relationship between total
luminosity and gas density to be superlinear. Consequently, the
resultant SFR-molecular line relation has index of order
unity.\label{figure:lighttodens}}
\end{figure}

We show this more quantitatively in Figure~\ref{figure:lighttodens},
where we plot the relative light (line flux) to density ratios versus
the mean cell density along a single sightline peering through the
nucleus of our sample disk galaxy. We plot the light to density ratios
for the CO (J=1-0) and CO (J=3-2) transitions, and normalize the
ratios at the highest density cell.  In
Figure~\ref{figure:lighttodens}, it is evident that the lower density
gas along the line of sight toward the nucleus proportionally emits
substantially more CO (J=3-2) intensity as a function of cell density
than than the thermalized gas \footnote{It is important to note that
there is no ``extra'' photon production from subthermally excited
gas. The emission from this diffuse gas is simply redistributed light
which largely originated in thermalized cores. It is the increased
relative light to density ratio (Figure~\ref{figure:lighttodens}) in
this subthermally excited gas along the line of sight that causes the
superlinear $\beta$ index (Equation~\ref{eq:lmolmh2}).}. This
translates to the CO (J=3-2) intensity tracing the
sightline-integrated mean gas density superlinearly
(Figure~\ref{figure:lvmeandens_los}).

\paragraph{Ground State Transition} 
\label{section:hcn10}
We now briefly turn our attention toward the SFR-HCN (J=1-0) relation
as it represents an instructive combination of the phenomena driving
the SFR-CO (J=1-0) and SFR-CO (J=3-2) relations.  At its root, the
characteristic emission pattern from HCN (J=1-0) falls into a similar
category as that of CO (J=3-2). That is to say, it is dominated by
subthermally excited gas, and the emission from these cells traces the
gas density in a superlinear manner
(e.g. Figure~\ref{figure:lighttodens}, which was shown for CO J=3-2,
but the results are applicable to HCN J=1-0 as well). Indeed, the
result is expected as the critical densities from the two lines are
only an order of magnitude different, and both are substantially
higher than the mean density of most clouds in our model galaxies. The
difference between HCN (J=1-0) emission and the CO (J=3-2) emission is
that there is no flat portion of the line intensity-gas density curve
as it is a ground state transition (Figure~\ref{figure:lcovmeandens},
right panel). The superlinear emission continues monotonically with
cell density across the full range of densities in the example galaxy
owing to nearly constant J=0 level populations and rising J=1 level
populations (Equation~\ref{eq:sourcefunction}). This was explicitly
seen in Figure~\ref{figure:poplevels} for the case of CO (though
qualitatively the case of HCN is similar). The intensity rises
superlinearly with increasing gas density because the excitation
processes to the J=1 rotational state have a large contribution from
the molecular line radiation field. The source of excitations from
line trapping increases superlinearly because the source of radiation
for the excitation - dense, thermalized cores - increases
superlinearly with $\bar{n}$.  This superlinear relation between \lmol
\ and $\bar{n}$ in the emitting gas cells results in a nearly linear
relationship between the SFR and HCN (J=1-0) luminosity in our models,
consistent with the well established observational results of
\citet{gao04a,gao04b}.

We therefore arrive at the following conclusions driving observed
molecular line-SFR relations:

\begin{itemize}

\item For lines with critical densities well below the mean density of
the clouds in the galaxy, the emission line will trace the total
molecular content of the galaxy. In these cases, we find an
SFR-molecular line luminosity index equivalent to the Schmidt-law
index.  This results in an SFR-\lcojone \ slope of $\sim$1.5 when the
SFR is constrained by SFR$\propto \rho^{1.5}$. Observationally, the
SFR-CO (J=1-0) index is found to lie between 1.4-1.6.

\item For lines with high critical densities, a superlinear relation
exists in \lmol \ and $\bar{n}$ in the gas owing to thermalized gas
lying only on the high density tail of the density distribution. The
light from the thermalized cores is redistributed such that the final
surface of emission is diffuse, subthermally excited gas along the
line of sight.  When the line is a ground state transition (e.g. HCN
J=1-0), the intensity from gas cells rises monotonically with mean
cloud density, though superlinearly owing to heavy contribution to the
excitation from line trapping. When the line is a transition above the
ground state (e.g. CO J=3-2), emission from the subthermally excited
cells is roughly constant with increasing mean gas density until the
level populations involved in the transition begin to approach LTE, at
which point the intensity is roughly linear with mean cloud
density. {\it In either case, for high critical density lines, a
superlinear increase in thermalized gas and high critical density
photon production with mean gas density results in an SFR-line
luminosity index ($\alpha$) lower than the Schmidt-law index.} In the
example of CO (J=3-2) and HCN (J=1-0) presented here, this results in
SFR-line luminosity indices of $\sim$1, consistent with the
measurements of \citet{nar05} and \citet{gao04a,gao04b}.

\end{itemize}

\subsection{Implications of Results}
\label{section:originsimplications}

The results presented thus far are a natural solution to the observed
SFR-CO and SFR-HCN relations.  Without any special tuning of
parameters, the same simulations are seen to additionally reproduce
characteristic CO emission line morphologies, intensities, excitation
conditions and effective radii of local disk galaxies and mergers
\citep{nar06a,nar07b}, as well as quasars at \zsim 6
\citep{nar07a}. More broadly, the same galaxy evolution simulations
have shown successes in reproducing a large body of characteristic
observable features of starburst galaxies, ULIRGs and quasars from
z=0-6 (\S~\ref{section:hydrodynamics}, and references therein).  In
this sense, the modeled reproduction of the observed SFR-CO \ and
SFR-HCN \ relations are a natural result of a series of simulations
which have reproduced many observed phenomena.

Second, we re-emphasize that the relationship between SFR and CO/HCN
emission is {\it built in} to the excitation mechanisms of both
diffuse and dense gas in galaxies, though should be taken in an
ensemble sense. Because the average physical conditions of the disk
galaxies and merger snapshots in our simulations result in molecular
line-gas density relations compatible with the observed SFR-molecular
line relations, no particular combination of disk galaxies and/or
merger snapshots was necessary for the reproduction of observed
relations in Figure~\ref{figure:sfrco}.

A natural question is whether or not any given SFR-molecular
luminosity relation holds particular significance as a physical SFR
``law'' relating the SFR to a property of the gas itself. The results
in this section show that all SFR-molecular line relations are
reflective of the underlying Schmidt law relating the star formation
rate to gas density. That is, our models suggest that the observed
linear SFR-molecular line luminosity relations (for high critical
density tracers) do not represent a fundamental indicator of the SFR,
but are rather simply indicative of the underlying Schmidt
law. Because we assumed that the SFR$\propto \rho^{1.5}$, the
resultant SFR-molecular line relations from the models matched the
observed relations rather well
(Equations~\ref{eq:sfrmh2}-\ref{eq:lmolmh2}). To some degree, these
models suggest that the observed SFR-CO \ and SFR-HCN \ relations
reflect a physical SFR law similar to the ones assumed for this model
(with index $N \approx$ 1.5).  Moreover, simulations have shown that
this choice of an SFR law reproduces the observed surface density
\citet{ken98a,ken98b} SFR laws well \citep{cox06c,spr00}. In this
sense, the predicted molecular SFR relations in this section are
simply reflective of the existing surface density SFR laws, as well as
volumetric gas density SFR laws.

Finally, we note that a general consequence of our models is that at
the highest SFRs, the SFR-\lmol \ slope will naturally turn toward the
underlying Schmidt index. This is because at these SFRs, the mean
density of the galaxies is typically high enough that the mean density
is of order the critical density of the molecular transition. In these
cases, the molecular line emission will be essentially counting clouds
in a manner similar to the CO (J=1-0) emission described earlier, and
an SFR-\lmol \ index of $\sim$1.5 will result. This is similar to the
interpretation advocated by \citet{krumthom07}.

\section{Testable Predictions}
\label{section:predictions}

\begin{figure*}
\includegraphics[angle=90,scale=0.35]{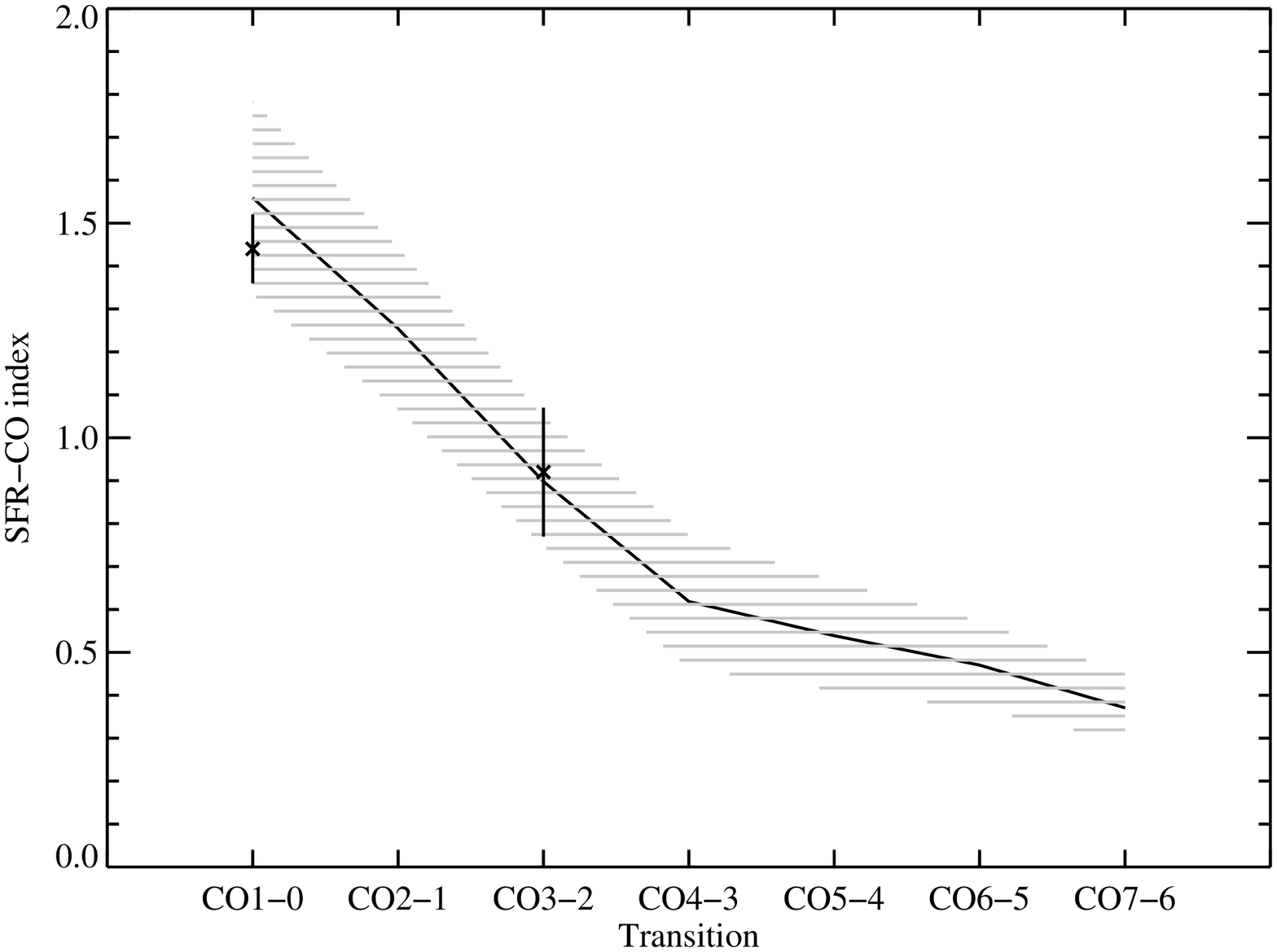}
\includegraphics[angle=90,scale=0.35]{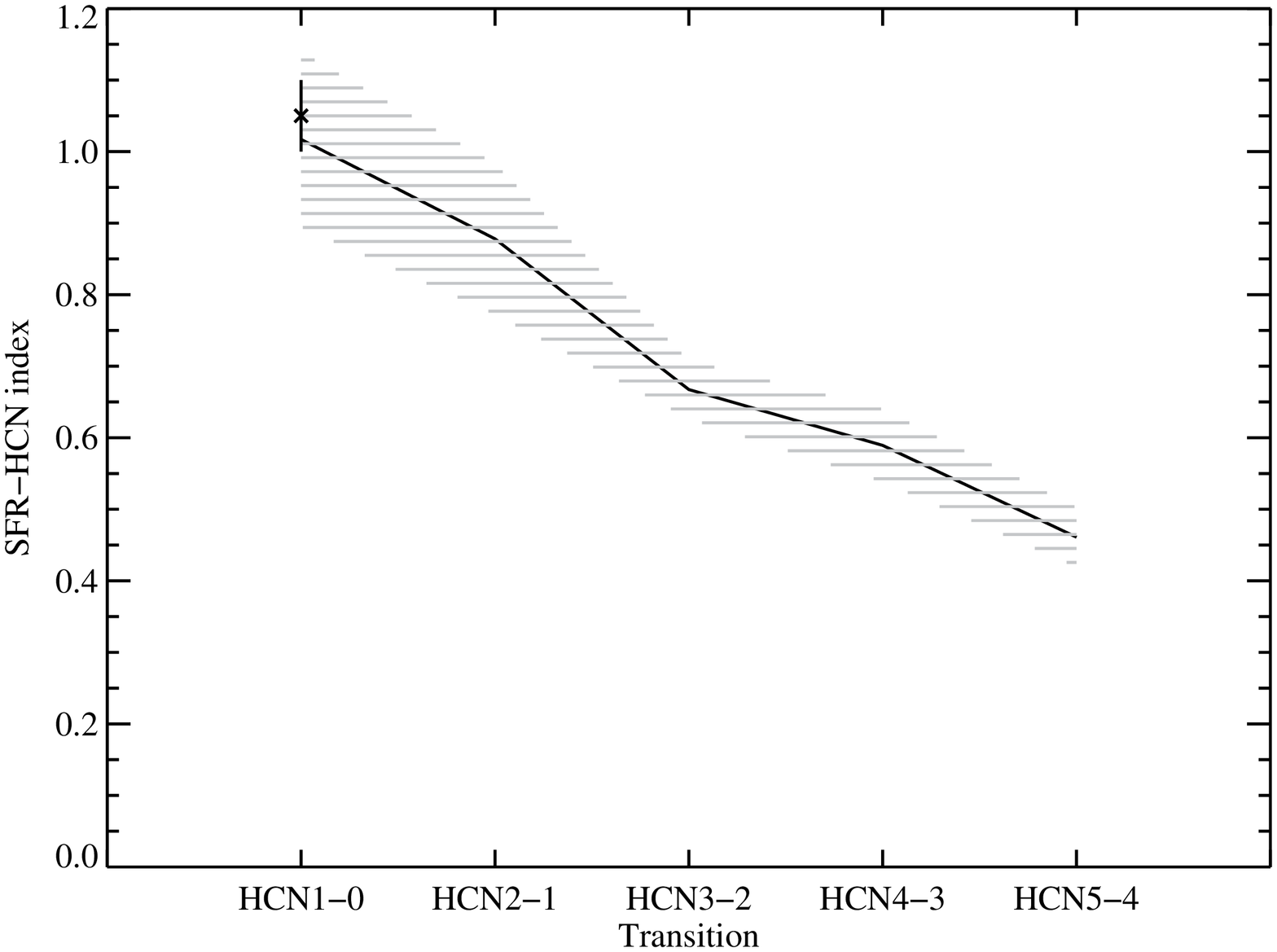}
\caption{ Predicted indices between SFR-CO and SFR-HCN relations for a
series of molecular line transitions. The SFR-CO (J=1-0), SFR-CO
(J=3-2) and SFR-HCN (J=1-0) indices are all consistent with the
observations of \citet{gao04a,gao04b} and \citet{nar05}. To simulate
observational variance with particular galaxy samples, we randomly
sampled 35 of our $\sim$100 model galaxies 100 times. The solid line
shows the mean derived slopes, and the shaded region the 1$\sigma$
contours in the dispersion of slopes. Additionally plotted are the
observed data from the surveys of \citet{gao04b} and \citet{nar05}
with their quoted error bars. These predicted slopes serve as a direct
observable test for these models.\label{figure:mvtransition}}
\end{figure*}

In Figure~\ref{figure:sfrco}, we showed for a random drawing of our
sample of $\sim$100 disk galaxies and merger snapshots that our model
results displayed consistent SFR-CO and SFR-HCN indices with
observations of local galaxies \citep{gao04a,gao04b,nar05,yao03}. We
can further extend these model results to make testable predictions
for the indices of unobserved SFR-CO and SFR-HCN transitions.

We plot the predicted indices for the range of readily observable
SFR-CO and SFR-HCN transitions in Figure~\ref{figure:mvtransition}. To
derive these results, we randomly drew 35 of our $\sim$100 galaxies
and took the best fitting slope between the SFR and molecular line
luminosity from the entire galaxy. Because we are modeling unresolved
observations of entire galaxies in our simulation sample, these
predictions can be directly applied to observations.  We did this for
each transition 100 times, and in Figure~\ref{figure:mvtransition}
denote the mean of these results with the solid line, and the standard
deviation in the dispersion with the hatched region. We additionally
plot the observed data from the surveys of \citet{gao04b} and
\citet{nar05} with associated error bars. The model results are quite
consistent with the three observational data points. Moreover, the
dispersion in the prediction is consistent with the resultant
dispersion in slopes of a random sampling of a comparable fraction of
galaxies in the \citet{gao04a,gao04b} sample.

The slope of the SFR-CO (J=1-0) relation roughly follows the assumed
Schmidt law as the CO (J=1-0) serves as an accurate tracer of the
total molecular gas. Higher lying transitions become shallower as the
relationship between the transition and the gas density (globally)
becomes superlinear. At the highest transitions
(e.g. CO $\ga$ 4), the decrease in the SFR-\lco \ slope begins to
flatten. Broadly, this owes to a relative plateau in the critical
densities as a function of increasing rotational transition.

{\it The SFR-\lmol \ index for tracers with critical density higher
  than HCN (J=1-0) or CO (J=3-2) is directly predicted to be
  sublinear}. This owes to the increasingly small fraction of gas in
galaxies which is thermalized in these transitions.  As the mean
density of galaxies increases, the amount of gas thermalized in these
transitions (in the most extreme peaks of the density distribution;
e.g. Figure~\ref{figure:meandens_ncrit}) increases
superlinearly. Alternatively said, the fraction of emission dominated
by subthermal emission in galaxies increases as the critical density
of the tracer increases. Consequently, the \lmol-$\bar{n}$ transition
becomes increasingly superlinear.

The predictions seen in Figure~\ref{figure:mvtransition} may serve as
model-distinguishing observational tests. That is, the standard
interpretation of the linear relationship seen between \lir \ and HCN
(J=1-0) luminosity and CO (J=3-2) luminosity is that the molecular
line emission traces dense gas more physically involved in the star
formation process (e.g. $n \ga$10$^{4-5}$\cmthree). In that picture,
observations of HCN and CO transitions with even higher critical
densities ought to similarly show a linear relationship between \lir \
and molecular line luminosity as they continue to probe dense star
forming cores \citep[e.g.][]{gao04a,gao04b}. In contrast, our models
suggest that the observed relations between SFR and molecular line
luminosity are driven by a superlinear increase in \lmol \ with
increasing $\bar{n}$, and will consequently have a sublinear SFR-\lmol
slope for higher critical density lines than CO (J=3-2) or HCN (J=1-0)
(Figure~\ref{figure:mvtransition}).  Observations of alternative CO
lines (e.g. CO J=2-1) or higher lying CO or HCN transitions in
galaxies will provide a direct test of these models.

We note that while the models of \citet{krumthom07} do not
explicitly predict the SFR-\lmol \ index for transitions beyond the
ground state, because the physical reasoning for their derived
SFR-\lmol \ relation for GMCs is similar to that predicted here, in
principle similar results can be expected. We discuss this work in the
context of their models more in \S~\ref{section:othertheory}.

The dependence of the SFR-\lmol \ relations on the relationship
between $\bar{n}$ and $n_{\rm crit}$ lends two other natural tests of
the models to be borne out. First, galaxies with the highest SFRs
typically have the largest mean densities. For these galaxies, because
$\bar{n} \sim n_{\rm crit}$, an SFR-\lmol \ index equal to the
underlying Schmidt index (here, 1.5) is to be expected. That is, the
slope of the SFR-\lmol \ relation should steepen for high critical
density tracers at the highest SFRs. We remind the reader that the
SFR-\lmol \ relations typically use \lir \ as a proxy for the SFR,
which may contain contamination from embedded AGNs (which we have not
included in these models). Thus, salient point of these models, as
well as those of \citet{krumthom07} regarding these high luminosity
points is that {\it regardless of the potential contribution of
  embedded AGNs} these dense, star forming systems will systematically
lie above the linear relation between \lir \ and HCN (J=1-0)/CO
(J=3-2) seen for lower luminosity (\lir $\la$ 10$^{12}$\lsunend)
systems.

Second, regions {\it within} galaxies with higher mean critical
density will have a different SFR-\lmol \ index than global
measurements. Lines of sight which probe the dense nucleus of an
e.g. ULIRG may have densities approaching that of the critical density
of high critical density tracers. In these cases, a superlinear
SFR-\lmol \ relation may be found.

\section{Comparison with Observations}
\label{section:observations}

The model results presented here quantitatively reproduce the observed
SFR-CO (J=1-0), SFR-CO (J=3-2), and SFR-HCN (J=1-0) indices for local
galaxies. These are found to be reflective of the ensemble-averaged
excitation conditions in these galaxies, and the manner in which the
molecular line luminosity is related to mean cloud density.

Surveys of Galactic GMC cores have shown that high dipole moment
molecules such as CS and HCN serve as an accurate tracer of dense gas
mass \citep[e.g.][]{shi03,wu05}. However, the interpretation from
extragalactic sources is mixed. Varying results have come from the
observational literature regarding the origin of traditional tracers
of dense gas (such as HCN) from extragalactic surveys. For example,
\citet{gre06} find the HCN emission in some systems to arise from
dense, thermalized cores, whereas \citet{pap07a} cite Arp 193 as an
example in which the bulk of the HCN emission arises from subthermally
populated gas. More broadly, constraints on high lying transitions in
CO, HCO$^+$ and HCN from the observational samples of \citet{gre06},
\citet{nar05}, \citet{pap07a}, \citet{pap07b} and \citet{yao03} among
others evidence a wide range of molecular excitation conditions.  In
this regard, the models presented here are consistent with these
observational results.

Sensitive observations of high redshift galaxies are beginning to
measure sources in the early Universe in terms of their place on the
SFR-molecular line relation as well \citep[e.g][ for an extensive
review, see Solomon \& Vanden Bout 2005 and references
therein]{gao07,gre05,hai06,rie06,wag05,wag07,wei07}. Owing to its
relatively high abundance, most detections at high-\z \ have been in
CO (with a smattering of HCN, HCO$^+$ and other molecules), and thus
we focus our comparisons with high-\z \ galaxies on CO.  A summary of
a nearly-current sample may be found in Figure 5 of \citet{rie06}. 

In general, the interpretation of SFR-molecular line relations at high
redshift are muddied by two factors. First, the increasing fraction of
AGN in high redshift infrared luminous sources almost certainly
contaminates the \lir \ from the observed galaxies and quasars, thus
causing a potentially significant overestimate in the SFR. Second, as
the objects span a large range in redshifts, the rest frame
transitions observed are typically quite diverse, and thus not always
probing the same phase of gas. The cumulative \lir-CO luminosity (over
numerous transitions) index for sources from z=0$\sim$6 as reported by
\citet{rie06} is $\sim$1.4, similar to the local SFR-CO (J=1-0)
relation. Certainly, at lower luminosities (\lir
$\la$10$^{12}$\lsunend), the molecular line data is dominated by CO
(J=1-0) observations of local galaxies. The relationship between \lir
\ and CO luminosity for these sources then is consistent with the
results of our simulations if the \lir \ in the lower luminosity
sources is dominated by dust heating by O and B stars. At higher
luminosities (\lir $\ga$ 10$^{12}$ \lsunend) the sources are
characteristically high redshift quasars, submillimeter galaxies and
radio galaxies. While the observed CO transition in these sources is
typically greater than the ground state transition, the relationship
between \lir \ and CO luminosity retains an index of $\sim$1.4. In
contrast, our models (\S~\ref{section:origins}) would predict a slope
less than $\sim$1.4-1.6 if a Schmidt law index of 1.4-1.6 was valid
for these high redshift sources. One possible origin for this steeper
slope is a contribution to the \lir \ from embedded AGN.

Alternatively, the models presented here \citep[as well as those of
][]{krumthom07} suggest that high critical density tracers such as HCN
(J=1-0) and high lying CO transitions may show a natural upturn from
the SFR-line luminosity relation at the highest luminosities even
without the contribution of embedded AGN.
The galaxies at this end of the luminosity range are typically massive
starbursts and/or advanced mergers with significant amounts of dense
gas. As discussed previously, and shown in
Figure~\ref{figure:lcovmeandens} in the high mean gas density regime,
tracers of dense gas such as CO (J=3-2) and HCN (J=1-0) become
thermalized. The galaxies in our simulations with the highest SFRs
(typically recently coalesced mergers) contain a large amount of dense
gas, and thus most of their e.g. CO (J=3-2) emission arises from
thermalized cells. The luminosity from this gas rises linearly with
increasing cloud density ($\beta$), and the consequent SFR-molecular
line relation ($\alpha$) will be superlinear (nearly equivalent to the
Schmidt SFR index) for a sample of these galaxies. This is consistent
with the modeling results of \citet{krumthom07} who find a similar
upturn in their SFR-molecular line relations when the mean cloud
density is much greater than the critical density of the molecular
line, as well as the observational results of \citet{gao07}.

We reiterate caution, however, that the observed SFR is typically
derived from \lir \ measurements, and at these high infrared
luminosities, the \lir \ may have a non-negligible contribution from a
central AGN \citep[e.g.][]{kim02,tra01,vei02}. While in principle our
hydrodynamic simulations have the capability to investigate the
potential contribution from growing black holes, a full calculation of
the IR SED as well as relating it to inferred SFR properties is
outside the scope of this work, and is deferred to a future study.

\section{Relationship to Other Models and Interpretations}
\label{section:othertheory}

We now turn our attention toward other models and interpretations for
the observed SFR-molecular line relations, and how our solution
compares to these works.

\citet{gao04b} interpret the tight linear correlation between \lir \
and HCN (J=1-0) luminosity as evidence for an increasing fraction of
dense gas in the most luminous sources in their sample, as well as a
constant star formation efficiency in terms of dense ($n \ga
$10$^{5}$\cmthree) molecular gas mass. Because stars form in the dense
cores of clouds, the linear relationship between \lir \ (which is
powered by star formation) and HCN (J=1-0) luminosity is interpreted
as a natural one.

 Our models find that the fraction of dense molecular gas naturally
 increases with star formation rate. This is true both for the mergers
 which funnel cold gas into the central kiloparsec (fueling starbursts
 of $\sim$100 \msunyrend), as well as isolated disk galaxies. Indeed
 this is an expected result as the SFR in our simulations is
 parametrized in terms of the cold gas density. This results in a
 higher fraction of the gas being thermalized in high critical density
 tracers in the systems with the highest SFRs (and, by extension,
 highest infrared luminosities). Indeed, this is what causes the
 upturn in the SFR-\lmol \ relation for high critical density tracers
 at the highest SFRs.

Utilizing observations of Galactic cloud cores, \citet{wu05} found
that the dense Galactic cloud cores showed a linear relationship
between \lir \ and HCN (J=1-0) emission. Using these results, they
posited that the linear relationship between \lir \ and HCN (J=1-0)
emission arises from HCN tracing fundamental star forming units which
scale self-similarly from star forming regions in the Galaxy to
ULIRGs. In the context that even high critical density tracers such as
HCN and CS are thermalized in star forming cores \citep[a result
supported by both observational and theoretical evidence,
e.g.][]{shi03,wal94b,wal94a}, the model results presented here may
suggest that a Kennicutt-Schmidt index of $N$=1.5 may not apply to the
star formation rate of dense cores. The simulations here show that
when the mean cloud density approaches the critical density of a
molecular line, the line luminosity faithfully traces the cloud
density (e.g. Figure~\ref{figure:lcovmeandens}, left panel). Thus, in
the context of these models, the relationship between the SFR and line
luminosity in dense, thermalized Galactic cores is expected to take an
index of $N$=1.5 if this index controls the Kennicutt-Schmidt SFR law
for these objects. It may be, then, that in light of the linear
relation between \lir \ and \lhcnjone \ that a linear SFR law
(e.g. SFR$\sim \rho$) describes star formation in dense cores
\citep{wu05}.

The recent study by \citet{krumthom07} utilized radiative transfer
modeling coupled with physical models of GMCs (consistent with
turbulence-regulated star formation) to derive a motivation for the
observed SFR-molecular line relations.  These authors found that for
individual star forming clouds, the line luminosity from high critical
density tracers such as HCN (J=1-0) increased superlinearly with mean
gas density owing to a superlinear increase in the fraction of
thermalized gas with increasing mean cloud density. This results in a
linear relationship between SFR and HCN (J=1-0) luminosity. While the
simulations by \citet{krumthom07} were geared toward physical models
of GMCs (in contrast to the hydrodynamic models of galaxies and galaxy
mergers presented here), the physical mechanisms driving the SFR-\lmol
\ relation are the same in both cases. In our simulations, a small
fraction of the gas is typically thermalized in high critical density
tracers (recall Figure~\ref{figure:meandens_ncrit}). As the mean
density of the distribution of cloud densities increases (e.g. as a
galaxy achieves a higher SFR, typically through compression of gas in
the nuclear regions), the fraction of gas that is thermalized in high
critical density tracers moves toward the mean and increases
superlinearly. An observable consequence of this in our simulations is
the excitation of subthermal gas along the LOS, which is readily
testable in local galaxy samples. In this sense, it is quite appealing
that two different modeling techniques of systems of different size
scales (GMCs versus models for galaxies) find the same physical
mechanism driving the observed SFR-\lmol \ relation. In both cases,
the fundamental SFR relation is indeed the Schmidt relation. The model
results presented here may be directly tested as they predict that
tracers of higher critical density than HCN (J=1-0) or CO (J=3-2) will
be thermalized in a small fraction of the galaxy's gas mass, and show
a sublinear relationship between SFR and line luminosity
(Figure~\ref{figure:mvtransition}).

Finally, we note that we include constant molecular abundances
throughout our model galaxies, as a full chemical reaction network is
both outside the scope of this work and not feasible given our
spatial resolution limitations. In this sense, we are unable to
evaluate these models in terms of potential HCN chemistry in the
vicinity of a hard X-ray flux as has been argued by some authors.
\citep[e.g.][]{com07,gra06,lin06}. We do note, though, that our models
quantitatively reproduce the observed relations between SFR and CO
(J=1-0), CO (J=3-2) and HCN (J=1-0) emission in local galaxies while
utilizing constant fractional molecular abundances. This may imply
that potential chemistry-related effects have a negligible effect on
observed SFR-molecular line relations.

\section{Conclusions and Summary}
\label{section:conclusions}

We have utilized a combination of 3D non-LTE radiative transfer
calculations with hydrodynamic simulations of isolated disk galaxies
and galaxy mergers to derive a physical model for the observed
SFR-molecular line relations. We specifically focus on the examples of
the SFR-CO (J=1-0), CO (J=3-2) and HCN (J=1-0) relations as they are
the best constrained observationally, and show that our model
quantitatively reproduces the observed relations when a Schmidt index
of $\sim$1.5 is assumed.

While the linear relationship between SFR and high critical density
tracers such as HCN (J=1-0) and CO (J=3-2) in galaxies have been
standardly interpreted as a fundamental SFR law owing to dense gas
being the formation site of massive stars, our models suggest that
this is not the entire story.  The linear relations of SFR and HCN
(J=1-0) and CO (J=3-2) arise as a consequence of a superlinear
relation between line luminosity and mean gas density in
galaxies. This owes to small amounts of thermalized gas in high
critical density tracers. The fundamental
relation is instead the underlying Schmidt law which sets the way in
which observed transitions trace the molecular gas.

Our model makes the prediction that for CO lines with J$_{\rm upper}
>$3 and HCN lines with J$_{\rm upper} >$2, the SFR-line luminosity
relationships will be sublinear
(Figure~\ref{figure:mvtransition}). These predictions can directly be
tested with existing submillimeter-wave technology, as well as with
ALMA.  Our models additionally provide specific interpretation
regarding the existing observed SFR-molecular line relations:

\begin{enumerate}

\item The slope in a given SFR-molecular line luminosity relation is
dependent on both the underlying Schmidt law controlling the SFR for
the galaxy and the relationship between molecular line luminosity and
density of emitting gas (the details of which are outlined in
\S~\ref{section:origins}). When line luminosity traces gas density
linearly, the resultant SFR-line luminosity index is similar to the
Schmidt law index. In cases where the line luminosity increases with
gas density superlinearly, the SFR is related to line luminosity with
an index less than the Schmidt index. The relationship between line
luminosity and gas density depends on how the critical density of the
line compares with the mean density of the bulk of the emitting
clouds. This is similar to the explanations posited by
\citet{krumthom07} who utilized models for turbulence-regulated
GMCs. This directly affects the observed relations between SFR and CO
(J=1-0), CO (J=3-2) and HCN (J=1-0) in the following way:

\begin{enumerate}
  \item Owing to its low critical density, the CO (J=1-0) line is
  roughly thermalized throughout most regions of the galaxies in our
  simulation sample. This results in a linear rise in CO (J=1-0)
  luminosity with increasing gas density, and a consequent SFR-CO
  (J=1-0) relationship with index similar to the Schmidt index. For
  the case of a Schmidt index of $\sim$1.5, our simulations reproduce
  the observed relation between SFR and CO (J=1-0) luminosity.

  \item The critical densities of CO (J=3=2) and HCN (J=1-0) are much
  higher than the mean density of the bulk of the clouds in our
  simulated galaxies. Because of this, the relationship between \lmol \
  and $\bar{n}$ is superlinear, causing an SFR-\lmol \ relation which has
  index less than the Schmidt index.  An important observational
  consequence of this is that significant amounts of emission can
  arise from subthermally excited diffuse gas in the vicinity of
  denser regions.  For the case of a Schmidt index of $\sim$1.5, the
  observed relations between SFR and HCN (J=1-0) and CO (J=3-2)
  emission are recovered.

\end{enumerate}

\item The emission processes driving the line luminosity-density
  relations (and consequently SFR-line luminosity relations) are
  variable such that some galaxies exhibit mostly thermalized gas for
  high critical density tracers whereas others are largely
  subthermally excited. Generally, the galaxies with higher SFRs have
  more of their gas thermalized which may drive the upturn in the
  \lir-HCN (J=1-0) relationship for extremely high luminosity sources
  observed by \citet{gao07}. This result is additionally recovered by
  models for GMCs by \citet{krumthom07}.

\item The physical basis for these models has a similar reasoning as
those posited by \citet{krumthom07} for the SFR-\lmol \ relation in
turbulence regulated GMCs. In this sense, two completely different
modeling techniques arrive at similar physical motivations for the
observed SFR-\lmol \ relation in galaxies.
 
\item We reemphasize that these models can be {\it directly tested}
via observations of high critical density tracers (see
Figure~\ref{figure:mvtransition}). In particular, these models suggest
that observations of HCN lines higher than J=1-0 and CO lines higher
than CO J=3-2 should bear sublinear relations between SFR and \lmol.

\item A natural consequence of this model is that at the highest SFRs,
  the SFR-\lmol relation will have an index similar to that of the
  underlying Schmidt index. This is a generic feature of our models,
  and happens regardless of the inclusion of AGN.

\end{enumerate}


\acknowledgements We thank Mark Krumholz and Todd Thompson for
detailed comments on a previous version of this text. We are grateful
to R. Shane Bussmann, Sukanya Chakrabarti, Kristian Finlator, Brandon
Kelly, Rob Kennicutt, Craig Kulesa, Yuexing Li, Chris Martin, Dominik
Riechers, Nick Scoville and A. Sternberg for conversations which
helped progress this work. Conversations with Joop Schaye helped
clarify much of the presentation in this work. We additionally thank
Christine Wilson for alerting us to an error in an early draft of this
work.  D.N. was supported by an NSF Graduate Research Fellowship
during part of this study. Support for this work was provided in part
by the Keck Foundation.  Support for this work was also provided by
NASA through grant number HST-AR-10308 from the Space Telescope
Science Institute, which is operated by AURA, Inc. under NASA contract
NAS5-26555. The calculations were performed in part at the Center for
Parallel Astrophysical Computing at the Harvard-Smithsonian Center for
Astrophysics.

\end{document}